\documentclass[12pt,ethesis]{thesis05}

\usepackage{url}

\newcommand{\ie}{\emph{i.e.,}}
\newcommand{\eg}{\emph{e.g.,}}

\usepackage{changebar}
\usepackage{latexsym}
\usepackage{fancyheadings}
\usepackage{multirow}
\usepackage{lscape}
\usepackage{amstext}
\usepackage{amsmath}
\usepackage{amssymb}
\usepackage{calc}
\usepackage{longtable}
\usepackage{algorithmic}
\usepackage{algorithm}
\usepackage{graphicx}

\newtheorem{thm}{Theorem}[section]
\newcommand{\ket}[1]{| #1 \rangle}
\newcommand{\bra}[1]{\langle #1 |}
\newcommand{\braket}[2]{\langle #1 | #2 \rangle}
\newcommand{\ketbra}[2]{| #1 \rangle\langle #2 |}

\newcommand{\cnot}{\mbox{C-NOT}}

\renewcommand{\bar}[1]{\overline{#1}}
\newenvironment{proof}{\noindent {\bf Proof : }}{$\qed$}

\newcommand{\G}{\mathcal{G}}
\renewcommand{\O}{O}
\newcommand{\qed}{\hspace{5mm}\square\vspace{4mm}}

\renewcommand{\perp}{\bot}

\newtheorem{theorem}{Theorem}

\newtheorem{lemma}[theorem]{Lemma}
\newtheorem{claim}[theorem]{Claim}
\newtheorem{cor}[theorem]{Corollary}
\newtheorem{fact}[theorem]{Fact}

\newtheorem{defn}[theorem]{Definition}
\newcommand{\comment}[1]{}
\newcommand{\suppress}[1]{}

\newcommand{\complex}{{\mathbb C}}

\title{Quantum Algorithm for Commutativity Testing of a Matrix Set}
\author{Yuki Kelly Itakura}
\discipline{Computer Science}
\degree{Master of Mathematics}

\newif\ifpdf
\ifx\pdfoutput\undefined
\pdffalse 
\else
\pdfoutput=1 
\pdftrue
\fi
\ifpdf
\usepackage[pdftex]{graphicx}
\else
\usepackage{graphicx}
\fi

\begin{document}

%

\ifpdf \DeclareGraphicsExtensions{.pdf, .jpg, .tif} \else
\DeclareGraphicsExtensions{.eps, .jpg} \fi

\prepages

\maketitle
\sigpages

\begin{abstract}
Suppose we have $k$ matrices of size $n \times n$. We are given an
oracle that knows all the entries of $k$ matrices, that is, we can
query the oracle an $(i,j)$ entry of the $l$-th matrix. The goal
is to test if each pair of $k$ matrices commute with each other or
not with as few queries to the oracle as possible. In order to
solve this problem, we use a theorem of Mario Szegedy~\cite{Sz04a,
Sz04b} that relates a hitting time of a classical random walk to
that of a quantum walk. We also take a look at another method of
quantum walk by Andris Ambainis~\cite{Amb03b}. We apply both walks
into triangle finding problem~\cite{MSS05} and matrix verification
problem~\cite{BS04} to compare the powers of the two different
walks. We also present Ambainis's method of lower bounding
technique~\cite{Amb00} to obtain a lower bound for this problem.
It turns out Szegedy's algorithm can be generalized to solve
similar problems. Therefore we use Szegedy's theorem to analyze
the problem of matrix set commutativity. We give an
$O(k^{4/5}n^{9/5})$ algorithm as well as a lower bound of
$\Omega(k^{1/2}n)$. We generalize the technique used in coming up
with the upper bound to solve a broader range of similar problems.
This is probably the first problem to be studied on the quantum
query complexity using quantum walks that involves more than one
parameter, here, $k$ and $n$.

\end{abstract}

\begin{acknowledgements}
The author would like to acknowledge Ashwin Nayak for supervion,
Richard Cleve for reading this essay, Andris Ambainis and Frederic
Magniez for consultation on lower bounds and the differences
between the two quantum walks respectively, as well as Mike Mosca
for the operation of IQC and Mike and Ophelia Lizaridis for the
funding of IQC.

The author would also like to acknowledge both her quantum and
classical friends, especially; Pierre Philipps for ``Tempest'',
Pranab Sen for regular helps, Alex Golynski for feeding her
brownies, the Crazy Lebanese Exchange Students (TM) for fun, and
all the people she danced with, including Scott Aaronson.

\end{acknowledgements}
\tableofcontents
\listoffigures
\listofalgorithms
\mainbody

\chapter{Introduction}\label{chap:intro}
\section{The Model, Motivation, and the Main
Results}\label{sec:intro} Suppose we are given a set $X$ of size
$n$ and we want to test if the set satisfies a given property. We
are also given an \emph{oracle} that computes $f(i)$ for some
index $i$ in the set. For example, in \emph{element
distinctness}~\cite{Amb03b}, $X$ is a set of integer variables,
$\{x_1, x_2, \ldots, x_n\}$ and the property to test is whether
there are two different indices $i$ and $j$ such that $x_i=x_j$.
In order to decide if $X$ satisfies the property, we query the
oracle for values $f(i)=x_i$ at various indices $i$. In general,
we are interested in minimizing the classical or quantum query
complexity, the number of queries a classical or quantum algorithm
make to the oracle. This notion will be defined formally in
Section~\ref{sec:G}.

We are interested in studying classical and quantum query
complexities because an oracle sometimes gives a separation
between them. For example, de Beaudrap, Cleve and Watrous showed
one problem where we need an exponentially many queries in the
bounded error classical case, but only a single query is needed in
the quantum case~\cite{dBCW02}. Another occasion to study a query
complexity is when obtaining a time complexity is hard. In such a
case, the number of queries we make gives a lower bound for the
time complexity. In fact, currently there is no lower bound method
for quantum time complexity that gives super-linear bounding, and
by studying quantum query complexity, we get lower bounds
heuristic on quantum time complexity.

One of the powers of quantum computation comes from the fact that
we can query in \emph{superposition}. That is, if we are given a
set of $n$ elements from $1$ to $n$ denoted $[n]$, we can query an
oracle in parallel \emph{once} to obtain a superposition of $f(1)$
through $f(n)$. However, as we will see in
Section~\ref{sec:superposition}, we can in a sense only learn one
of the $f(i)$'s from such a query. The real power of quantum
computation comes from \emph{interference}. That is, the
information in the states, \eg\ $f(i)$'s, can be combined by means
of unitary quantum gates in a non-trivial way, and we can extract
a global property of the inputs. For example, in Deutsch's
algorithm, given two input bits indexed by $0$ and $1$, we cannot
obtain both $f(0)$ and $f(1)$ in one oracle query. However, by
making a suitable quantum query, we can obtain a global property,
$f(0)\oplus f(1)$~\cite{Deu85}. This interference is also used for
quantum search in an unstructured database, in an algorithm due to
Grover~\cite{Gro98}, to extract a global property, \ie\ if the set
we are given contains an element we are looking for.

It turns out we can generalize Grover's search to test if a set we
have satisfies a given property using a quantum version of a
random walk, called a \emph{quantum walk}. Using a quantum walk,
for example, element distinctness can be solved in
$O(n^{2/3})$~\cite{Amb03b} queries with a matching lower bound of
$\Omega(n^{2/3})$~\cite{AS04}. A quantum walk was first studied on
the line, both discrete~\cite{ABN+01} and continuous~\cite{FG98},
analogous to classical discrete and continuous random walks,
except that a quantum discrete walk uses a \emph{coin} to decide
which point to move to next, whereas a quantum continuous walk
does not. The discrete quantum walk on the line showed that the
probability distribution after certain number of steps of quantum
walk is different from that of the classical probability
distribution~\cite{ABN+01}. The continuous quantum walk was then
applied to a graph that gave an exponential speed up in a hitting
time as compared to the classical counterpart~\cite{CFG02}. The
discrete quantum walk on the line was also extended to general
graphs~\cite{AAKV01} and later applied to a search on a
hypercube~\cite{SKW03}. Both discrete~\cite{AKR05} and
continuous~\cite{CG04} walks were applied to search an item on a
grid. Ambainis~\cite{Amb03b} used a discrete quantum walk to solve
element distinctness. This is generalized in~\cite{MSS05} to find
a three clique in a graph (\emph{triangle finding}). 
Szegedy proposed a different quantization of a classical Markov
chain in~\cite{Sz04a, Sz04b}. He showed that there is a
quadratic speedup for the hitting time of his quantization of classical walk. 
Szegedy's quantization was applied in~\cite{BS04} to verify a
product of two matrices (\emph{matrix verification}). 
For more details in the development of quantum walk based
algorithms, see~\cite{Amb04}.

The goal of this essay is to investigate the query complexity of
testing the commutativity of $k$ matrices of size $n\times n$.
This essay is probably the first to study quantum query complexity
that involves two variables, $k$, the number of matrices in the
set and $n$, their dimension. We show that there are three upper
bounds for this problem, $O(kn^{5/3})$, $O(k^{2/3}n^2)$ and
$O(k^{4/5}n^{9/5})$, depending on the relationships between the
variables $k$ and $n$. We also show a lower bound of
$\Omega(k^{1/2}n)$.

The organization of the essay is as follows. We first introduce
the mathematical background necessary to understand our quantum
algorithms in Section~\ref{sec:math-basics}. Then we take a look
at the details of Szegedy-Walk in Section~\ref{sec:S-walk} and
Ambainis-Walk in Section~\ref{sec:A-walk}. We use these two walks
to analyze triangle finding problem in Section~\ref{sec:triangle}
to see a case where Amabinis-Walk performs better. In
Section~\ref{sec:lowerbound}, we take a look at a quantum
adversary method~\cite{Amb00} to obtain a lower bound for our
problems. We shift our focus to matrices next and in
Section~\ref{sec:matrix-ver}, we study matrix verification
problem. In Chapter~\ref{chap:Matrix-testing}, we finally study
the problem of testing the commutativity of $k$ matrices of size
$n \times n$. We first take a look at a case where $k=2$ by using
a modification of matrix verification in
Section~\ref{sec:MTSinglePair}. Next we study four different
algorithms for a general $k$ in Section~\ref{sec:Commutativity}. 
This problem is generalized in
Section~\ref{sec:commutativity-general}. Finally we give a summary
and directions for future work in Chapter~\ref{chap:conclusions}.

\section{Mathematical Background}\label{sec:math-basics}
In this section, we will go over the mathematical background
necessary to follow the algorithms in this essay. Beyond the
content in this section,~\cite{NC00} is a good reference in
general introductory material in quantum computation.

\subsection{Space and qubit}\label{sec:qubits}
Classically, information is encoded in a binary string using a
sequence of bits $0$ and $1$.  Quantumly, information is encoded
in a finite-dimensional complex vector space, endowed with the
standard inner product, a Hilbert space using \emph{qubits}. A
qubit may exist in states $\ket{0}$ and $\ket{1}$, which are basis
vectors for a two-dimensional space. 
\[
\ket{0}=\left(
        \begin{array}{c}
        1\\
        0\\
        \end{array}
        \right)
\]
and
\[
\ket{1}=\left(
        \begin{array}{c}
        0\\
        1\\
        \end{array}
        \right),\]
or in any linear combination of these basis states with unit norm.
We call the two vectors $\ket{0}$ and $\ket{1}$
\emph{computational basis} for the two-dimensional Hilbert space
since they correspond to the conventional bit representation of
information. There are other pairs of basis states that span the
two-dimensional Hilbert space
but we focus on the computational basis. 
The state of a sequence of $n$ qubits is a unit vector in the
$n$-fold tensor product space $\complex^2 \otimes \complex^2
\otimes \ldots \otimes \complex^2$. This $2^n$-dimensional space
is spanned by tensor products of states $\ket{0}$, $\ket{1}$. This
is the computational basis for the $n$-qubit memory. The tensor
product of two vectors $\ket{\phi}$ and $\ket{\psi}$ is denoted as
$\ket{\phi} \otimes \ket{\psi}$. When these are computational
basis vectors given by bit strings~$x,y$, we may abbreviate the
state $\ket{x} \otimes \ket{y}$ by $\ket{x,y}$ or simply
$\ket{xy}$.
%
The latter two make sense when $x$ and $y$ are bit-strings. Using
a standard vector notation, a tensor product of two vectors is
obtained by multiplying each entry in the left vector with the
right vector.
\[
\left( \begin{array}{c}
      a\\
      b\\
      \end{array}
\right) \bigotimes \left( \begin{array}{c}
      c\\
      d\\
      e\\
      \end{array}
\right) = \left( \begin{array}{c}
      a \left( \begin{array}{c}
      c\\
      d\\
      e\\
      \end{array}\right)\\
      b\left( \begin{array}{c}
      c\\
      d\\
      e\\
      \end{array}\right)\\
      \end{array}
\right) = \left( \begin{array}{c}
      ac\\
      ad\\
      ae\\
      bc\\
      bd\\
      be\\
      \end{array}
\right) .\]

For example, a two qubit state $\ket{10}$ is,
\[
\ket{1}\ket{0} = \left(
        \begin{array}{c}
        0\\
        1\\
        \end{array}
        \right) \otimes
        \left(
        \begin{array}{c}
        1\\
        0\\
        \end{array}
        \right)
        =
        \left(
        \begin{array}{c}
        0\\
        0\\
        1\\
        0\\
        \end{array}
        \right).
\]
This extends in the natural way to tensor products of
higher-dimensional vectors.  The dual of the vector $\ket{i}$ is
denoted by $\bra{i}$, which is a row vector obtained by taking a
conjugate transpose of $\ket{i}$. For $\ket{0}$ this is just a row
vector $(1\; 0)$.

\subsection{Superposition and
Measurement}\label{sec:superposition} A Hilbert space of dimension
$n$ is spanned by $n$ orthonormal vectors, and we can express a
state in the space as a linear combination of these basis states.
For a two-dimensional Hilbert space with the basis states
$\ket{0}$ and $\ket{1}$, any state $\ket{\phi_1}$ can be expressed
as
\[
\ket{\phi_1}=\frac{1}{\sqrt{|\alpha_0|^2+|\alpha_1|^2}}\left(\alpha_0
\ket{0} + \alpha_1 \ket{1}\right),
\]
where $\alpha_i$ is the \emph{amplitude} of $\ket{i}$. Similarly,
a multiple qubit state is also expressed as a linear combination
of its basis states. For example, a two qubit state $\ket{\phi_2}$
can be expressed as a linear combination of four computational
basis states,
\[
\ket{\phi_2}=\frac{1}{\sqrt{|\alpha_{00}|^2+|\alpha_{01}|^2+|\alpha_{10}|^2+|\alpha_{11}|^2}}
\left(\alpha_{00} \ket{00} + \alpha_{01} \ket{01} + \alpha_{10}
\ket{10} + \alpha_{11} \ket{11}\right).
\]
\begin{defn}[Measurement~\cite{NC00}]
Given a set of $n$ basis states $\{\ket{m_i}\}$, a
\emph{measurement} in a basis $\ket{m}$ of a state
$\ket{\phi_n}=\alpha_1 \ket{m_1} + \ldots + \alpha_{n}
\ket{m_{n}}$ is a \emph{projection} of $\ket{\phi_n}$ onto one of
the basis states by applying projective operators
$\{\ketbra{m_i}{m_i}\}$ to $\ket{\phi_n}$. The superposition
\emph{collapses} to one of the basis states and the probability of
obtaining $\ket{m_i}$ is
$\bra{\phi_n}\left(\ketbra{m_i}{m_i}\right)\ket{\phi_n}=|\alpha_i|^2$.
The state after measurement is then,
$\frac{\ket{m_i}\braket{m_i}{\phi_n}}{|\alpha_i|}$.
\end{defn}
The implication above is that before measuring $\ket{\phi_n}$, the
state is in \emph{superposition} of its basis states, but
measuring collapses the superposition and gives only one of the
basis states as an outcome with the probability according to the
amplitude of the basis states in $\ket{\phi_n}$. Since the
probabilities must sum up to one, this means that the sum of the
squares of the amplitudes must also sum up to one,
\[
\displaystyle\sum_{i=1}^{n} | \alpha_i |^2 =1.
\]

Also note that we \emph{normalize} the collapsed state resulting
from the measurement so that the squares of the amplitudes in this
new state also sums up to one.

For a multiple qubit system, we can also measure a small set of
qubits only and leave the rest alone. A measurement in the
computational basis of the first qubit collapses the first qubit
into one outcome of the measurement, the remaining state is
unchanged. Formally, the state is projected onto a subspace
consistent to the measurement outcome. For example, if we have
\[
\ket{\psi}=\frac{1}{\sqrt{2}}(\ket{00}+\ket{01}+\ket{10}+\ket{11}),
\]
measuring the first qubit gives $0$ with probability $\frac{1}{2}$
and $1$ with probability $\frac{1}{2}$. On outcome $0$,  the new
state is
\[
\ket{\psi'}=\frac{1}{\sqrt{2}}(\ket{00}+\ket{01}),
\]
and on outcome $1$, the new state is
\[
\ket{\psi''}=\frac{1}{\sqrt{2}}(\ket{10}+\ket{11}).
\]
\subsection{Operators and Quantum Gates}\label{sec:operators}
A quantum gate is a matrix that acts on the state vectors. In
order for a matrix to be a legal (physically realizable)
operator, it must be \emph{unitary}, that is $U^{\dag}U=I$, where
$U^{\dag}$ is the
conjugate transpose of a gate $U$. 
%
Some gates that are used for the construction of the algorithms in
this essay are X, Hadamard $H$, and control-NOT gates.
\[
X=\left( \begin{array}{cc}
        0 & 1 \\
        1 & 0 \\
        \end{array}
  \right),
H=\frac{1}{\sqrt{2}}\left( \begin{array}{cc}
        1 & 1 \\
        1 & -1 \\
        \end{array}
  \right),
\cnot=\left( \begin{array}{cccc}
        1 & 0 & 0& 0 \\
        0 & 1 & 0 & 0 \\
        0 & 0 & 0 & 1 \\
        0 & 0 & 1 & 0 \\
        \end{array}
  \right)
\]
\begin{figure}
  \begin{center}
\includegraphics[height=2.5in,width=3.5in,angle=0]{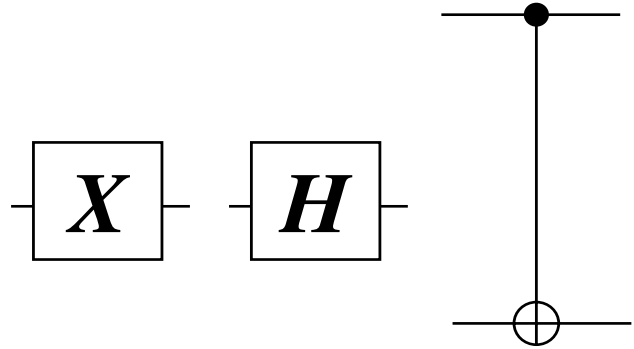}
 \end{center}
  \caption{Diagrammatic Representations of $X$, $H$, and control-NOT respectively.}
  \label{fig:gates}
\end{figure}
The effect of $X$ on computational basis is a logical NOT
operation, $X\ket{0}=\ket{1}$ and $X\ket{1}=\ket{0}$. A Hadamard,
$H$ transforms $\ket{0}$ into a uniform superposition of $\ket{0}$
and $\ket{1}$ \ie\ $\frac{(\ket{0}+\ket{1})}{\sqrt{2}}$ and
$\ket{1}$ into $\frac{(\ket{0}-\ket{1})}{\sqrt{2}}$. Applying $H$
for each of $n$ qubits initialized to $\ket{0}$, we can create a
uniform superposition of $2^n$ computational bases, \ie\
$\frac{1}{\sqrt{2^n}}\displaystyle\sum_{x=0}^{{2^n}-1}\ket{x}$.
C-NOT takes two qubits as inputs and conditioned on the first
qubit, it performs a logical NOT operation to the second qubit,
\eg\ $\cnot \ket{01}=\ket{01}$ because the first qubit is $0$, and
$\cnot \ket{11}=\ket{10}$ because the first qubit is $1$. It is a
unitary operation corresponding to a classical gate.

Recall that classically if we are given NOT and AND gates, we can
construct a classical circuit for any boolean function. Such a set
of gates is called a \emph{universal set} of gates. Similarly,
quantumly, we have universal sets of gates. This means that any
unitary transformation on $n$ quantum bits maybe approximated to
within a specified $\epsilon>1$ (in the spectral norm, say) by
composing a sequence of these gates. One example involves the use
of a C-NOT and a Hadamard with two additional one-qubit gates
called a \emph{phase} gate $S$, and $\pi/8$ gate $T$.
\[
S=\left( \begin{array}{cc}
        1 & 0 \\
        0 & i \\
        \end{array}
  \right),
T=\left( \begin{array}{cc}
        1 & 0 \\
        0 & e^{i\pi/4} \\
        \end{array}
  \right)
\]

For the proof of the universality of this gate set, refer Section
4.5 in~\cite{NC00}.
\subsection{Quantum Algorithms and the Circuit Model}
\label{sec:circuit}  
\begin{figure}
  \begin{center}
\includegraphics[height=2.5in,width=3.5in,angle=0]{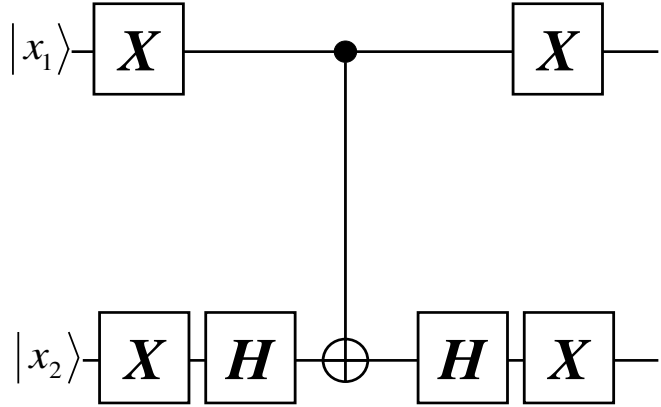}
 \end{center}
  \caption{A Circuit that Implements a Phase Flip}
  \label{fig:circuit}
\end{figure}

A quantum algorithm consists of \emph{quantum
registers} that hold qubits, and a series of unitary operations
described by a \emph{quantum circuit}. The registers are
initialized to $\ket{0}$ except for the input register which is
initialized to the bits of the problem instance, as in a classical
circuit. The circuit consists of a sequence of gates from a
universal set of quantum gates with the labels of the qubits the
gates are applied to. In Figure~\ref{fig:circuit}, the registers
are represented by black lines. As we apply operators we move from
the left to the right of the circuit. At the end of the algorithm,
\ie\ at the right end of the circuit, a measurement is performed
on one or more qubits in the computational basis, which gives an
outcome of the algorithm. An algorithm is said to compute a
boolean function with \emph{bounded error} if when the input
string $x$ is in the language, the algorithm accepts $x$ (has
outcome $1$) with probability more than $3/4$, and when $x$ is not
in the language, the algorithm accepts (\ie\ has outcome $0$) with
probability less than $1/4$.

For example, suppose we want to implement an algorithm that
\emph{flips the phase} if the registers both contain $\ket{0}$,
but not otherwise. Then Figure~\ref{fig:circuit} performs such
algorithm. It first applies an $X$ gate to each register, and then
applies a Hadamard gate to the second qubit, followed by a C-NOT
conditioned on the first qubit, followed by a Hadamard on the
second qubit, and finally applies $X$ gates to two of the qubits.
This operator can be written as
\[
I-2\ketbra{00}{00}.
\]

It is straightforward to check that both the circuit and the above
matrix flips the phase of the qubit if they are both $0$.

The quantum time complexity of a (boolean) function is measured by
the least number of gates required to implement an algorithm that
computes the function with bounded error in terms of the size of
the input. In Figure~\ref{fig:circuit}, the input size is two and
the number of gates is seven.

\subsection{Query Model and Quantum Query Complexity}
\label{sec:G} We first formally define an oracle in terms of an
operator.
\begin{defn}[Oracle]~\cite{NC00}
An oracle $\O$ for a function $f:\{1, \ldots, n\}\rightarrow
\{0,1\}$ is a unitary operator that acts on a computational basis
such that
\[
\O\ket{x}\ket{q}=\ket{x}\ket{q\oplus f(x)},
\]
where $\ket{q}$ is an oracle qubit with $q \in \{0,1\}$, which is
flipped conditioned on $x\in \{1, \ldots, n\}$, \ie\ flipped if
$f(x)=1$. An oracle for a function with a larger range, $\{1,
\ldots, n\}$ is defined similarly, with $O(\log n)$ qubits each
for the query and the function value and $\oplus$ representing a
bit-wise XOR.
\end{defn}

Using an oracle, we can perform a query algorithm,
\begin{defn}[$T$-Query Quantum Algorithm]~\cite{BBCM+01}
A $T$-query quantum algorithm $A$ with an oracle $\O$ for function
$f$ is defined as
\[
A=U_T \O U_{T-1}\ldots \O U_1 \O U_0,
\]
where all the transformation are defined on a three register
quantum memory consisting of the query register, the oracle
response register and workplace qubits for the algorithms. The
$U_i$'s are unitary transformations independent of the function
$f$, and the algorithm only depends on the function $f$ through
$T$ applications of $\O$.
\end{defn}

The query complexity of an algorithm is measured by the number of
oracle operators we apply. The query complexity of computing a
property $P$ of the oracle function $f$ is given by the least
query complexity algorithm that computes $P(f)$.

For a search algorithm where an oracle outputs $f(x)=1$ if $x$ is
a target of the search and the property being if a given set
contains a target element, we usually prepare an oracle qubit as
$\frac{\ket{0}-\ket{1}}{\sqrt{2}}$, so that we get
\[
\O\frac{1}{\sqrt{2^n}}\displaystyle\sum_{x=0}^{2^n-1}\ket{x}\left(\frac{\ket{0}-\ket{1}}{\sqrt{2}}\right)
\rightarrow
(-1)^{f(x)}\frac{1}{\sqrt{2^n}}\displaystyle\sum_{x=0}^{2^n-1}\ket{x}\left(\frac{\ket{0}-\ket{1}}{\sqrt{2}}\right).
\]

Since the oracle qubit does not change throughout the algorithm,
we could simply think of this oracle as flipping a phase if
$f(x)=1$.

What would be the action of $\O$ in a search algorithm? Suppose we
have an initial state
\[
\ket{\psi}=\frac{1}{\sqrt{2^n}}\displaystyle\sum_{x=0}^{2^n-1}\ket{x},
\]
and that $\ket{\psi}$ is a combination of two vectors,
$\ket{\alpha}$ and $\ket{\beta}$, where the former is a uniform
superposition of elements $x$ such that $f(x)=0$, and the latter
contains the rest of elements. Then the act of applying the oracle
is a \emph{reflection} about the axis $\ket{\alpha}$ because
\[
\O(a\ket{\alpha}+b\ket{\beta})=a\ket{\alpha}-b\ket{\beta}.
\]

Recall the phase flip operator from the last section, which up to
an overall sign of $-1$ is a reflection operator. Thus we can
create the following reflection operator by removing $X$ gates in
Figure~\ref{fig:circuit}.
\[
2\ketbra{0}{0}-I.
\]

This construction extends in a straightforward manner to $n$
qubits. In general, in order to implement a phase flip on qubits
that represent $n$, $O(\log{n})$ gates are required.

We can create another reflection operator also called \emph{Grover
diffusion operator} that reflects the state with the axis
$\ket{\psi}$ by
\[
H^{\oplus n}(2\ketbra{0}{0}-I)H^{\oplus n}=2\ketbra{\psi}{\psi}-I.
\]

Hence so far we have two reflection operators, $\O$ and
$2\ketbra{\psi}{\psi}-I$.

\begin{lemma}\label{lemma:rotation}
Applying
\[
G=\left(2\ketbra{\psi}{\psi}-I\right)\O
\]
is a rotation in a two-dimensional space spanned by $\ket{\alpha}$
and $\ket{\beta}$ by $2\theta$, where $\theta$ is the initial
angle between $\ket{\psi}$ and $\ket{\alpha}$.
\end{lemma}

\begin{figure}
  \begin{center}
\includegraphics[height=3.5in,width=5in,angle=0]{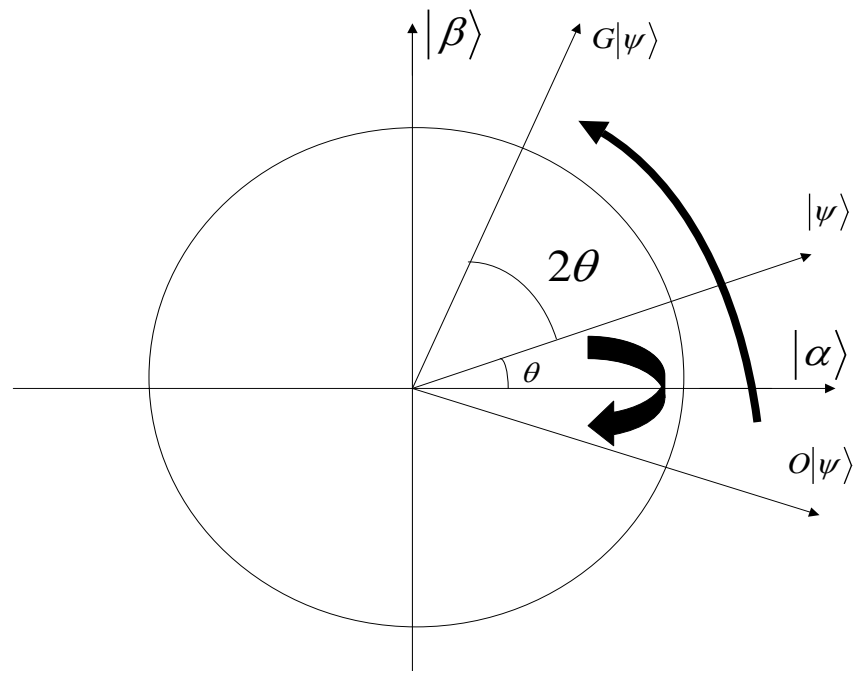}
  \end{center}
  \caption{A Product of Two Reflections is a Rotation}
  \label{fig:rotation}
\end{figure}

Lemma~\ref{lemma:rotation} also holds for the composition of
reflections of any two vectors. We will use this fact later in
this essay. In Figure~\ref{fig:rotation}, the action of $G$ is
described geometrically. It first reflects $\ket{\psi}$ about the
axis $\ket{\alpha}$, and then $O\ket{\psi}$ is reflected against
the original state $\ket{\psi}$.
In this one step of $G$, there is only one query $\O$. In Grover's
algorithm, this process is repeated $O(\sqrt{N})$ times for
$N=2^n$ so as to rotate the state of the query register close to
$\ket{\beta}$: in the worst case when there is only one $x$ such
that $f(x)=1$, $\theta \approx \frac{1}{\sqrt{n}}$. This gives a
query complexity of $O(\sqrt{n})$.

\subsection{Reducing Error Probability}\label{sec:error} In many quantum
algorithms, we encounter a problem of reducing the error
probability from a constant such as $1/4$ to polynomial close to
$0$. An algorithm is said to compute a function $f$ with one-sided
error given an input $x$ if the following two conditions hold,
\begin{enumerate}
 \item If $x$ is not in the language, it rejects with probability
$1$.
 \item if $x$ is in the language, it accepts with probability at
 least $\epsilon>0$.
\end{enumerate}
This means that we have a probability $(1-\epsilon)$ of having a
false negative. In order to reduce the error probability to at
most $1/2$, we repeat this algorithm for $k=\lceil -
\frac{1}{\log{(1-\epsilon)}}\rceil\approx \frac{1}{\epsilon}$
times, because
\[
(1-\epsilon)^{ - \frac{1}{\log{(1-\epsilon)}}} \leq \frac{1}{2}.
\]
During any one time in the $k$ repetition of the algorithms, if
the algorithm accepts $x$, we terminate and decide ``yes''. Most
of the algorithms in this essay have one-sided error. For example,
in element distinctness, if we find two different indices $i$ and
$j$ such that $x_i \neq x_j$, we are sure it is in fact true.

In our algorithms, we will often compose bounded error quantum
algorithms. In such cases, a quantum algorithm is used as a
subroutine in place of an oracle. We would have to amplify, by
repetition, the success probability of the subroutine so that the
overall algorithm succeeds. This results in an additional factor
of $O(\log T)$ in the query complexity where the complexity with
an ideal oracle is $O(T)$. Such a scenario is studied
in~\cite{HMdW03} as a quantum search with a bounded error oracle.
The main result in~\cite{HMdW03} is that we only need to invoke
the oracle $\sqrt{n}$ times as opposed to the obvious approach
that gives $\sqrt{n}\log{n}$.

In this essay, whenever we have a one-sided error and we wish to
amplify the success probability, we assume the procedure is
modified as above. Moreover, if we have a case of imperfect oracle
realized by a bounded error quantum algorithm, we apply the
theorem in~\cite{HMdW03}.

\chapter{Related Work}\label{chap:prev-work}
\section{Quantum Walk of Szegedy}\label{sec:S-walk}
\subsection{Element Distinctness}\label{sec:ED}
Recall from Chapter~\ref{chap:intro}, the problem of Element
Distinctness: given a function $f:[n]=\{1,2,\ldots, n\} \mapsto
[m]$, $m\geq n$, as an oracle, we want to test if $f$ is one-one
or not. If $f$ is not one-one, we say there is a \emph{collision}.
That is, $(i,j)$ collide if $f(i)=f(j)$, The function $f$ can also
be written as a list of numbers: $f\equiv (f_1, f_2, \ldots,
f_n)$. The goal of the algorithm is to answer this question with
as few queries to the oracle as possible.

The significance of this problem is that it is one of the
applications of quantum walks that gives better bounds than classical counterparts. 
Underlying this quantum algorithm is a random walk. Ambainis was
the first to adopt this classical walk into a quantum
algorithm~\cite{Amb04}. 
Classically, the straightforward algorithm to solve Element
Distinctness 
is to go through the list one by one. Interestingly, this
straightforward algorithm performs better than a random walk based
algorithm classically which we will see in
Section~\ref{sec:C-walk}.
\begin{fact}
Classical query complexity of element distinctness is  $\Theta
(n)$.
\end{fact}
Since it is optimal an unordered search may be reduced to element
distinctness. However, in quantum scenario, quantum walk based
algorithm performs better than the above bound. Quantum walk based
algorithm is a quantum version of a random walk based algorithm,
which is described below.
\subsection{Classical Walk Based Algorithm}\label{sec:C-walk}
The following is a classical algorithm based on random walk for
finding a collision.
\begin{algorithm}
\caption{A Classical Walk Algorithm for Element
Distinctness}\label{alg:ED}
\begin{algorithmic}[1]
 \STATE Pick a uniformly random set $I$ of $r$ elements out of
$[n]$ (call it an $r$-subset).
 \STATE Query $f$ at points in $I$.
 \IF{There is a collision within $I$}
  \RETURN {``Collision found'' and elements that collide}
 \ENDIF

 \COMMENT{Walk on $r$-subsets of $[n]$.}

 \COMMENT{(idea) Pick an element in the set and one not in the set uniformly
 at random (\emph{u.a.r.}). Swap
these elements. Note that we are maintaining the size of the
subset.}

 \FOR{ $ t \leq T$ }

   \STATE Pick $i \in I$ and $j \in [n]-I$ u.a.r.
   \STATE  $I \leftarrow (I-\{i\})\cup \{j\}$.

   \STATE Query $f_j$.
   \IF{There is a collision within $I$}
     \STATE Output the elements that collide.
     \RETURN
   \ENDIF
 \ENDFOR
 \PRINT{``No collision''}
 \COMMENT{\ie\ $f$ is one-one.}
\end{algorithmic}
\end{algorithm}
This walk is \emph{irreducible} in the sense that there is a path
between any pair of subsets. 
Let $\mathcal{T}$ be the first time the walk ``hits'' an
$r$-subset containing a collision (\emph{hitting time}).

\textbf{Observation}
\[
Pr(\text{walk stops in two steps $|$ any state}) \geq 1\cdot
\frac{2}{n-r}\cdot \frac{r-1}{r}\cdot \frac{1}{n-r} .\]
\begin{figure}
\includegraphics[height=2in,width=6in,angle=0]{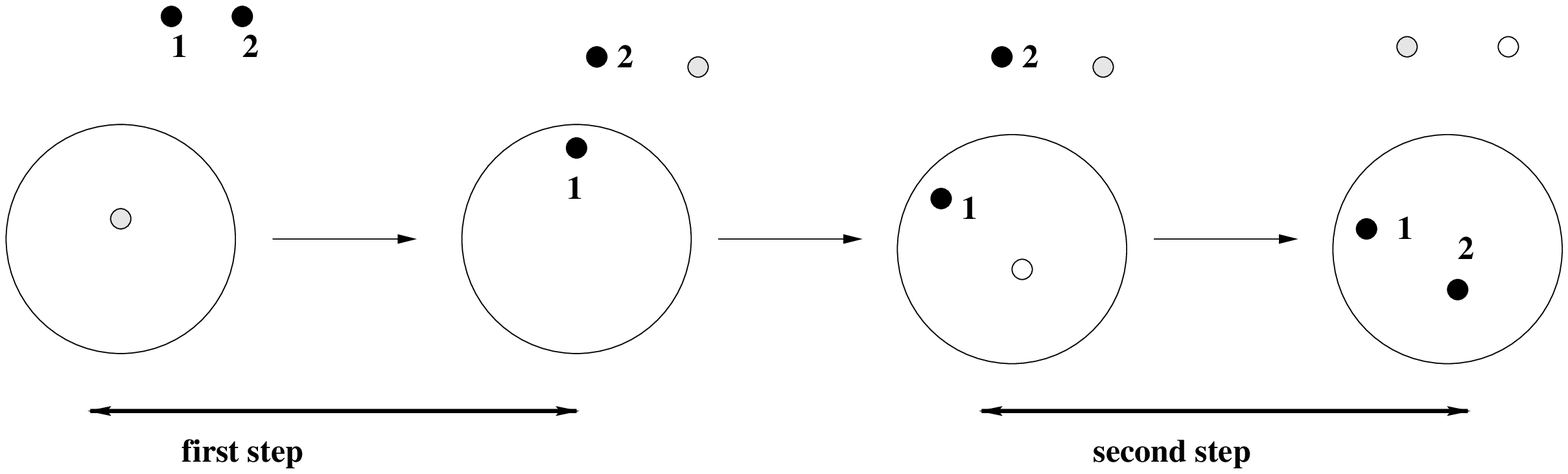}
  \caption{The Probability of the Walk Stopping in Two Steps}
  \label{fig:twoStepWalk}
\end{figure}

This is because in the worst case, there are exactly two elements
that collide with each other, and initially, we do not have any
element that form a colliding pair in the $r$-subset. Next we pick
one of the two colliding elements from $n-r$ elements not in the
set with probability $\frac{2}{n-r}$. In the second step, we first
choose an element in the $r$-subset that is not part of the
colliding set with probability $\frac{r-1}{r}$, and then we pick
the other colliding element not in the $r$-subset with probability
$\frac{1}{n-r}$ and swap these. In Figure~\ref{fig:twoStepWalk},
$1$ and $2$ are colliding elements not in the subset initially. It
describes a sequence of transformation by which they are found by
the algorithm. 
The hitting time for the walk is
\[
E(\mathcal{T}) \leq  \frac{(n-r)^2r}{2(r-1)}.\]

There are more sophisticated arguments that give a better bound on
$\mathcal{T}$. 
We will analyze hitting times more precisely.

\subsection{Hitting Time in Classical Walks}\label{sec:CHittingTime} Consider a Markov
Chain on the state space $X$, ($|X|=N$) given by the transition
matrix $P$, where $P=\left(p_{x,y}\right)$, $x, y \in X$ and
\[p_{x,y}=Pr(\text{making a transition to $y$ $|$ current state $=x$})
.\]

This corresponds to general Markov Chains, in the sense that if we
are at $x$, we move to any arbitrary state $y$ in the state space
with probability $p_{x,y}$. $P$ is called a stochastic matrix,
\ie\ $\displaystyle\sum_{y}p_{x,y}=1$ for all $x$. So all the rows
sum up to $1$.

We assume that the Markov Chain is
\begin{enumerate}
\item Symmetric: $p_{x,y}=p_{y,x}$. This makes the underlying
graph of the walk undirected.

\item Irreducible: There is a path between every pair of states.

\item Aperiodic:
 There exists $x\in X$ and $t_x
\geq 1$ such that
\[Pr(\text{We reach $x$ in $t$ steps starting from $x$})>0\]
for all $t \geq t_x$. Aperiodicity of the walk is equivalent to
having an underlying graph that is not bipartite. This implies the
same property for all $y \in X$ if the second property holds.
\end{enumerate}
\begin{figure}
  \begin{center}
\includegraphics[height=1.5in,width=1.5in,angle=0]{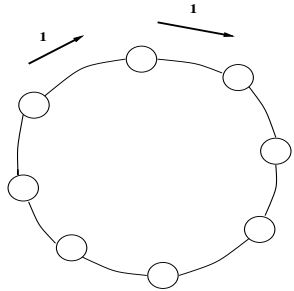}
  \end{center}
  \caption{An Example of a Periodic Markov Chain}
  \label{fig:periodicMC}
\end{figure}

What are the properties of such Markov Chains?
\begin{enumerate}
\item Since $P$ is symmetric, it is equal to its transpose,
$P=P^{\texttt{T}}$. So $P$ is doubly stochastic; both rows and
columns sum up to $1$.

\item Let $s$ be any initial distribution, then
$s^{\texttt{T}}P^{t}\longmapsto \pi^{\texttt{T}}=(\frac{1}{N},
\frac{1}{N},\ldots, \frac{1}{N})$ as $t\rightarrow \infty $ in the
$l_1$ metric. The distribution $\pi$ is called \emph{stationary
distribution}, which is a fixed point in a Markov Chain. So we
have uniform stationary distribution.

\item $\pi^{\texttt{T}}P=\pi^{\texttt{T}}=(\frac{1}{N},
\frac{1}{N},\ldots, \frac{1}{N})$: $\pi$ is an eigenvector of $P$
with eigenvalue of $1$. Since $P$ is symmetric, it is Hermitian,
therefore it is diagonalizable and all the eigenvalues are real.
Moreover, the other eigenvalues are strictly less than $1$:
$\lambda_1=1 > \lambda_2 \geq \ldots \geq \lambda_n > -1$. The
eigenvalue of $1$ is obtained from the irreducibility property.
Aperiodicity implies all the eigenvalues are $> -1$.
\end{enumerate}

In general, a \emph{marked state} $M \subseteq X$ is a subset
. For element distinctness, $M$ contains two
colliding elements. Since we stop at a marked state, the
transition matrix for this state is different from others. Suppose
we would like to search for one of the marked states by simulating
the walk and stopping when we see a state $x \in M$. The
transition matrix now looks like
\begin{equation}\label{eq:pm}
\tilde {P_M}=
 \left( \begin{array}{ccc}
P_M &P'  \\
0 & I  \\
\end{array} \right),
\end{equation}
where $(P_M \; P')$ are the rows of $P$ corresponding to $X-M$,
$P_M$ is $P$ from which rows and columns corresponding to $M$ have
been removed. The rows corresponding to the states in $M$ are $(0
\; I)$ since once we reach $M$, we do not move to any other state.

What is the hitting time of $M$? Let $T$ be the hitting time for
finding a marked state starting in distribution $s$.
\begin{fact}
\label{fact:hitting}
\begin{displaymath}
E(T)=s_{M}^{\texttt{T}}(I-P_{M})^{-1}\cdot \bold{1},
\end{displaymath}
\end{fact}
where $s_M$ is the projection of $s$ onto $X-M$, and
$\bold{1}^{\texttt{T}}=(1, 1, \ldots, 1)$. When $M$ is non-empty,
and since the Markov Chain is ergodic all the eigenvalues of $P_M$
have absolute value less than $1$. Therefore the expression is
well-defined.

\begin{proof} For any non-negative integer-valued random variable $T$,
$E(T)=\displaystyle\sum_{t=0}^{\infty}Pr(T
>t)$. In our case, $Pr(T>t)$ is the probability we have not reached the
marked state after $t$ steps. This is also the probability that we
are still in one of the states in $X-M$. Since the state
distribution after $t$ steps is $s^{\texttt{T}}\tilde {P}^{t}$,
where
\[\tilde {P}^{t}=
 \left( \begin{array}{ccc}
P_M^t & P'(t)  \\
0 & I  \\
\end{array} \right).
\]

Let $\bold{1}_{X-M}$ denote a vector that contains $1$ for the
first $|X-M|$ entries and $0$ for the rest. Then we have
\[
Pr(\text{We are not in a marked state after $t$
steps})=s^{T}\tilde{P}^{t} 1_{X-M}
=s^{T}_M\tilde{P}^{t}_{M}\bold{1}.
\]

Then,
\[
\begin{array}{ll}
E(T)& =\displaystyle\sum_{t=0}^{\infty}s_M^TP_M^t\bold{1}\\
\\
&=s_M^T(\displaystyle\sum_{t=0}^{\infty}P_M^t)\bold{1}\\
\\
&=s_M^T(I-P_M)^{-1}\bold{1}.\\
\end{array}
\]
\end{proof}
%
%
%

Stationary distribution for $P$ is a uniform distribution over all
elements. Thus by judicially choosing initial state to be the
stationary distribution, we get a good bound on hitting time.
\begin{cor}\label{cor:cHtime}
a. If $s=(\frac{1}{N}, \frac{1}{N}, \ldots, \frac{1}{N})$, then
hitting time $E(T)$ is
\[
E(T)=\frac{1}{N}\cdot \bold{1}\cdot (I-P_M)^{-1}\cdot \bold{1}.
\]

b. Let $\bold{1}_M=\frac{(1,\ldots, 1)}{\sqrt{N}}$. If the
eigenvalues/vectors of $P_M$ are $(\lambda_i , v_i)$ and
$\bold{1}_M=\displaystyle\sum_{i=1}^{N-m}\nu_i v_i$ then,
\begin{equation}
\label{eq:et} E(T)=\displaystyle\sum_{i=1}^{N-m}\nu_i^2
(\frac{1}{1-\lambda_i}),
\end{equation}
where $N$ is the normalization factor, $m=|M|$ is the size of
marked subsets, and $\lambda_i$ is the $i$-th largest eigenvalue
of $P_M$ in magnitude.
\end{cor}

Note in the first part of Corollary~\ref{cor:cHtime}, the
eigenvalues of $(I-P_M)^{-1}$ are $\frac{1}{1-\lambda_i}$, for
each eigenvalue $\lambda_i$ of $P_M$. Also since we are working
with real symmetric matrices, all the numbers $\nu_i$ are real.

The matrix $P_M$ is real, all the absolute values of eigenvalues
of $P_M$ are strictly less than $1$, and along with the symmetry
of $P_M$, it is orthogonally diagonalizable. This means we can
choose ${v_i}$ such that they form an orthonormal set.
Note that the spectral norm of a matrix is the largest singular
value of the matrix. Since $P_M$ is symmetric, it is equivalent to
the largest eigenvalue. Hence $\| P_M \| = \lambda_1$. Since
$\displaystyle\sum \nu_i^2$ is at most $1$, we have, $E(T) \leq
\frac{1}{1-\lambda_1}= \frac{1}{1-\| P_M \|}$.

In order to bound the hitting time, then we need to bound the
largest eigenvalue of $P_M$. 
\begin{lemma}[\cite{Sz04a}]
If the spectral gap($=1-\lambda_2(P)$) of $P$ is $\geq \delta$,
and if $\frac{|M|}{|X|}\geq \epsilon$, then $\| P_M \| \leq 1-
\frac{\delta \epsilon}{2}$. \label{lm:delta-epsilon}
\end{lemma}

In the above lemma, we define $\lambda_i(P)$ to be $i$-th largest
eigenvalue of $P$ in magnitude. Note that since $P$ has a uniform
distribution, $\lambda_1=1$. So the spectral gap, which is
formally, the difference between the largest and the second
largest eigenvalues in magnitude, is
$\lambda_1(P)-\lambda_2(P)=1-\lambda_2(P)$.
\begin{figure}
\begin{center}
\includegraphics[height=2in,width=3in,angle=0]{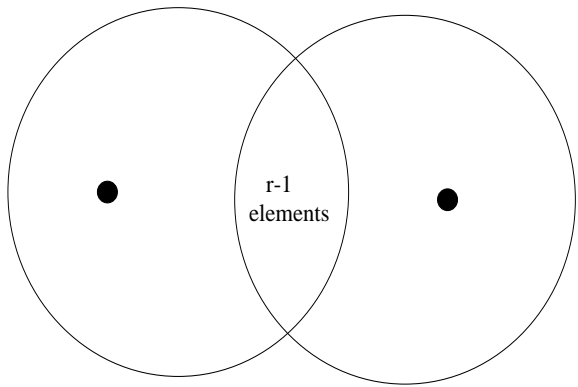}
  \end{center}
  \caption{$P_{ij}$ Moves from One State to Another
  With a Symmetric Difference of Two}
  \label{fig:AmazingMatrix}
\end{figure}

To bound the hitting time of the walk , we would like an explicit
formula for the spectral gap of $P$ to compute the upper bound of
the spectra for $P_M$. Recall 
the state space of the walk is $X=\{\text{$r$-subsets of }[n]\}$.
Given an $r$-subset, there are $r(n-r)$ other $r$-subsets to
transition to by swapping one of the $r$ elements in the current
subset with one of the $n-r$ elements not in the subset. Each of
these $r(n-r)$ subsets have equal probability of being moved to
from the current $r$-subset. Then
\[
 N=|X|={
n \choose r},
\]
and
\[
\begin{array}{ll}
p_{i,j}&=\left\{
\begin{array}{cc}
\frac{1}{r(n-r)} & \textrm{ if $|i \cap j|=r-1$}\\
0\\
\end{array}
\right.\\
& =\frac{J_{n,r,r-1}}{r(n-r)} ,
\end{array}
\]
where $J_{n,r,r-1}$ is a boolean matrix with entry $1$ iff $i$ and
$j$ are subsets of size $r$, whose intersection is of size $r-1$.
\begin{theorem}[\cite{Knu91}]
There are $r+1$ eigenspaces of $J_{n,r,r-1}$, eigenvalues
corresponding to
\[
\lambda_j=(r-j)(n-r)-j(r-j+1),\; 0 \leq j\leq r.
\]
\label{thm:Knu}
\end{theorem}
We have $r \leq \frac{n}{2}$, otherwise we have a high probability
of solving the problem in Line 3 in Algorithm~\ref{alg:ED}. Also,
$\lambda_j$ is a decreasing function of $j$. The eigenvalues are
not all positive, \eg\ for $j=r$, we have $\lambda_r=-r$. However,
we are only interested in the first and the second largest
eigenvalues, which are, $\lambda_0=r(n-r)$ and
$\lambda_1=r(n-r)-n$. Since these are eigenvalues for
$J_{n,r,r-1}$ and $P=\frac{J_{n,r,r-1}}{r(n-r)}$, the second
largest eigenvalue for $P$ is $\frac{r(n-r)-n}{r(n-r)}$. From
these, we can compute the spectral gap:
$1-\frac{\lambda_1}{r(n-r)}=\frac{n}{r(n-r)} > \frac{1}{r}$.
Remembering that $M$ is the set of $r$-subsets that contain a
colliding pair of elements, in order to lower bound the fraction
of marked elements, we need to consider the worst case scenario
where we have exactly one pair of colliding elements.
\[
\frac{|M|}{|X|} \geq \frac{ {n \choose r-2} } { { n \choose r } }
=\frac{r(r-1)}{(n-r-2)(n-r-1)} \geq \frac{r^2}{2n^2}
\]
for $r=o(n)$ when approximation involved.

From this, we have
\[
\begin{array}{ll}
\| P_M \| &\leq 1- \frac{\frac{r^2}{2n^2}\frac{1}{r}}
{2}\\
&=1-\frac{r}{4n^2}.\\
\end{array}
 \] So
 \[E(T) \leq \frac{1}{1- \| P_m \|} \leq
\frac{4n^2}{r}=O\left(\frac{n^2}{r}\right) .\]

This is a bound on the hitting time of the algorithm. The query
complexity of the algorithm is calculated as follows. We need to
make $r$ initial queries for the values of each element in the
initial $r$-subset. At each of the $O(\frac{n^2}{r})$ iteration of
the walk, we need to query the value of the new element we swapped
into the subset. Thus, we have $O(r+\frac{n^2}{r})$ query
complexity. This is minimized when $r=n$ and gives $O(n)$ query
complexity. This is equivalent to checking every element in the
entire set, thus giving no speedup to the straightforward
algorithm of sequentially checking every element in the set.

Now we are interested in the quantization of the classical
algorithm we have discussed thus far.

\subsection{Quantization of the classical
walk}\label{sec:quantization}
\begin{figure}
  \begin{center}
\includegraphics[height=2.5in,width=3in,angle=0]{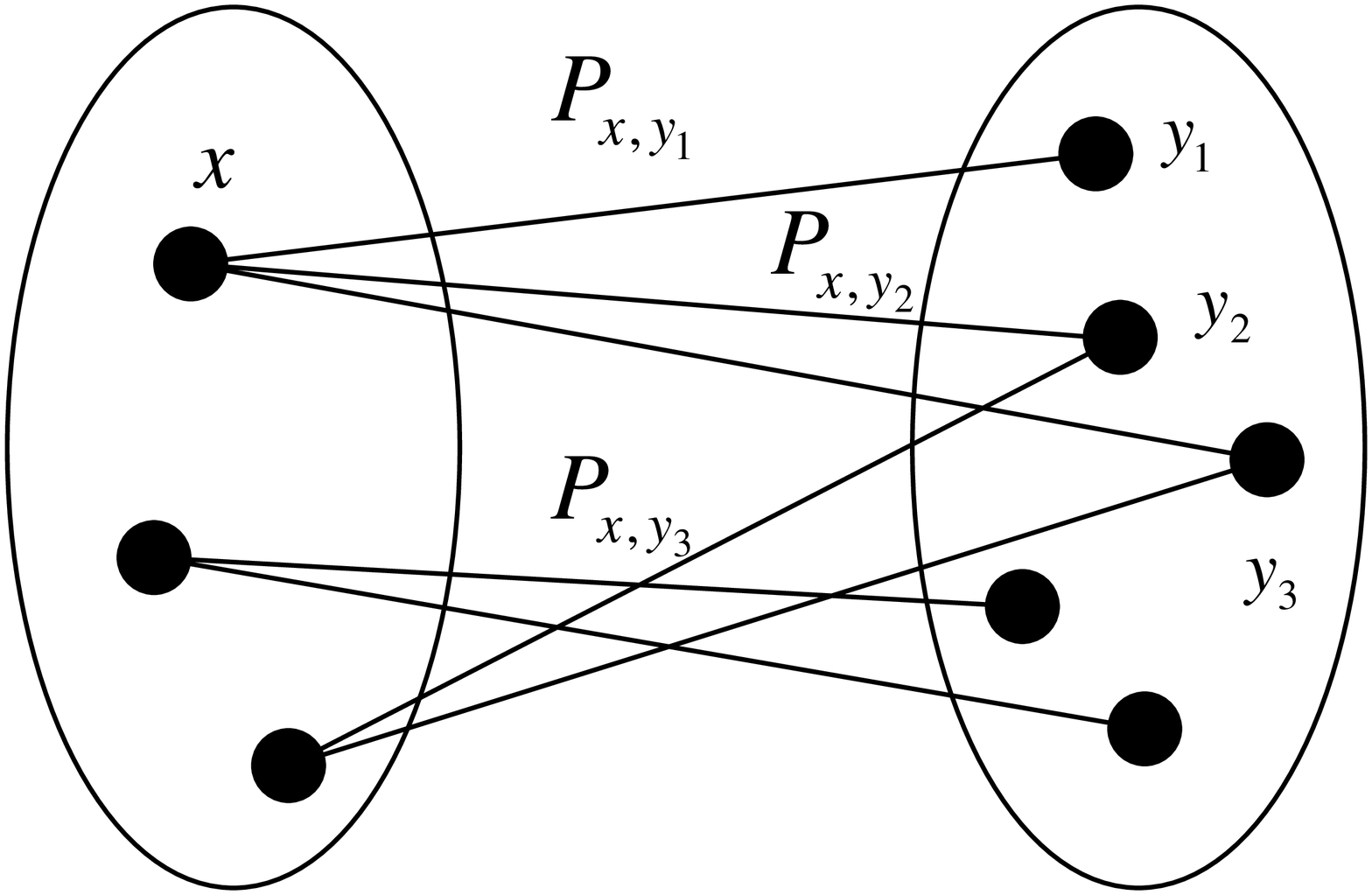}
  \end{center}
  \caption{A Bipartite Walk}
  \label{fig:bipartiteWalk}
\end{figure}

The first quantization of random walk in Algorithm~\ref{alg:ED}
was proposed by Ambainis~\cite{Amb04}, which is described in
Section~\ref{sec:A-walk}. A new kind of quantization of
classical walks was proposed by Szegedy~\cite{Sz04a}, which we present here in detail. 
The walk is over a bipartite graph. Each side of the graph
contains $r$-subsets as vertices. A pair of vertices in the left
and the right hand side of the graph are connected only if one can
be changed into another on the opposite side by removing one of
the elements in the subset and adding one that is not in the
current set. This is equivalent to having two vertices connected
if they differ in exactly two elements. The probability of moving
from a subset $x$ in the left side of the graph to a subset $y$ in
the right side of the graph is given by $p_{x,y}$. For each side
of the graph, we create a state,
\[\ket{\phi_x}=\displaystyle\sum_{y}\sqrt{p_{x,y}}\ket{x}\ket{y}\]
for the transition from $x$ on the left side to all of its
neighbors $y$ on the right side of the graph, and
\[\ket{\psi_y}=\displaystyle\sum_{x}\sqrt{p_{x,y}}\ket{x}\ket{y}\]
similarly. Note that,

a. $\{\phi_x\}_x$ and $\{\psi_y\}_y$ are orthonormal sets because
each $\ket{x}$ and $\ket{y}$ are distinct.

b. Let $E_1=span\{\phi_x\}_x$ and $E_2=span\{\psi_y\}_y$ and
$\pi_1$, $\pi_2$ to be orthonormal projections onto $E_1$, $E_2$
respectively. We define two unitary operators $R_1$ and $R_2$ as
$R_1=2\Pi_1-I$ and $R_2=2\Pi_2-I$.

Then $R_1$ is unitary because it can be implemented using the
combination of unitary gates similar to the reflection operator in
Section~\ref{sec:circuit}. We can see that $R_1$ is actually a
reflection operator about the space $E_1$ because
\[
\begin{array}{ll}
R_1\ket{\varphi} &= (2\pi_1 - I)\ket{\varphi}\\
                &= (2 \displaystyle\sum_x \ketbra{\phi_x}{\phi_x}-I)\ket{\varphi} \\
                &= \displaystyle\sum_x(2  \ketbra{\phi_x}{\phi_x}-I)\ket{\varphi} \\
                &= \left\{\begin{array}{cc}
                   \ket{\varphi} & \text{if $\varphi\in E_1$} \\
                   -\ket{\varphi} & \text{if $\varphi\in E_1^{\perp}$}\\
                   \end{array} \right.\\
\end{array}
\]
Similarly $R_2$ is unitary and is a reflection about the space
$E_2$.
\begin{defn}[Quantitization of a M.C. $P$]
\begin{displaymath}
W_P=R_2R_1
\end{displaymath}
\end{defn}

Why is this a natural definition? The straightforward way to
define a step of the walk is
\[
\begin{array}{ll}
\ket{x}\ket{\bar{0}} &\rightarrow
\ket{x}\displaystyle\displaystyle\sum_{y}\sqrt{p_{x,y}}\ket{y}\\
&\rightarrow
\ket{x\displaystyle}\displaystyle\sum_{y}\sqrt{p_{x,y}}\ket{y}\displaystyle\displaystyle\sum_{z}\sqrt{p_{y,z}}\ket{z}\\
&\rightarrow
\ket{x}\displaystyle\displaystyle\sum_{y}\sqrt{p_{x,y}}\ket{y}\displaystyle\displaystyle\sum_{z}\sqrt{p_{y,z}}\ket{z}\displaystyle\displaystyle\sum_{a}\sqrt{p_{z,a}}\ket{a}\\
\end{array}
\]

But this is just a simulation of a classical walk. 
Instead, we keep the memory of the previous step only. To do so,
we need an operator that diffuses $x$ into all the neighbors $y$
and vice versa.

Another way to see that the definition of the quantized walk is
natural is to look at the Grover diffusion operator as an operator
to move from a vertex in a complete graph to one of the adjacent
vertices with equal probability. This idea was introduced
in~\cite{Wat01}. Algorithm~\ref{alg:S-Walk} describes Szegedy's
algorithm.
\begin{algorithm}
\caption{Szegedy's Quantization of a Random Walk}
\label{alg:S-Walk}
\begin{algorithmic}[1]
 \STATE Let
 $\ket{\phi_0}=\frac{1}{\sqrt{N}}\displaystyle\sum_{x}\ket{\phi_x}=
 \frac{1}{\sqrt{N}}\displaystyle\sum_{x,y}\sqrt{p_{x,y}}\ket{x}\ket{y}=\frac{1}{\sqrt{N}}\displaystyle\sum_{y}\ket{\phi_y}$

 \STATE Measure if the first register is marked or not.
 \IF{The first register is marked}
  \RETURN \texttt{``Found marked element.''}
 \ENDIF

 \STATE Apply $H \otimes I$ to $\ket{0}\ket{\phi_{01}}$ to get
 $\frac{1}{\sqrt{2}}(\ket{0}\ket{\phi_{01}}+\ket{1}\ket{\phi_{01}})$

 \STATE Pick $t$ u.a.r. from $[0,\ldots, T]$

   \FOR{$i \leq T$}
    {
    \STATE Apply controlled-$W_{\tilde{P}}$ conditioned on the
    first register.

    \COMMENT{The modified matrix $\tilde{P}$ in Equation~\ref{eq:pm} lets us remain in the same state once we
    arrive at the marked state.}
    }
    \ENDFOR

 \COMMENT{ Now we have
 $\frac{1}{\sqrt{2}}(\ket{0}\ket{\phi_{01}}+\ket{1}W_{\tilde{P}}^{t}\ket{\phi_{01}})$}

 \STATE Apply $H \otimes I$ to get
     $\frac{1}{2}\ket{0}(\ket{\phi_{01}}+W_{\tilde{P}}^{t}\ket{\phi_{01}})+\frac{1}{2}\ket{1}(\ket{\phi_{01}}-W_{\tilde{P}}^{t}\ket{\phi_{01}})$

 \STATE Measure the first register.
   \IF{There is a '1' in the first register}
     \RETURN \texttt{``Detected marked element.''}
   \ELSE
     \RETURN \texttt{``No marked element.''}
   \ENDIF
\end{algorithmic}
\end{algorithm}


Szegedy defines the quantum hitting time as follows.
\begin{defn}[\cite{Sz04b}]\label{def:deviationTime}
$T$ is an $c$-deviation-on-average time with respect to
$\ket{\phi}$ if
\[
\frac{1}{T+1}\displaystyle\sum_{t=0}^{T}\|
W_{p}^{t}\ket{\phi}-\ket{\phi}\|_{2}^{2} \geq c _.
\]
\end{defn}

This was defined so that after $T$ steps of the walk, the average
deviation of the initial state is very high. That is, the state is
significantly skewed towards the marked state and so the
probability of observing the marked state is high. Since the walk
is realized by unitary evolution it cannot end up in a marked
state. Instead, it can cycle through states with high amplitude on
marked states.

Next we compute the complexity of this algorithm. One step of the
walk is $R_1$ followed by $R_2$. We show here that $R_1$ can be
implemented efficiently. $R_2$ can be implemented similarly.
Recall that
\[
\begin{array}{ll}
R_1&=(2\Pi_1-I)\\
&=\left( 2 \displaystyle\displaystyle\sum_{x}\ketbra{\phi_x}{\phi_x}-I \right)\\
&=\displaystyle\displaystyle\sum_{x}\left( 2\ketbra{\phi_x}{\phi_x}-I_x \right),\\
\end{array}
\]
where $I_x$ is identity on $\ket{x} \otimes \mathbb{C}^x$. The
last line follows from the fact that we are working on
\[
\begin{array}{ll}
\mathbb{C}^{X \times X}&\cong \mathbb{C}^X \otimes
\mathbb{C}^X\\
&\cong \bigoplus_x \ket{x} \otimes \mathbb{C}^X ,\\
\end{array}
\]
so $R_1$ which acts on $\mathbb{C}^X \otimes \mathbb{C}^X$ can be
decomposed into the direct sum of $|X|$ diffusion operators,
$2\ketbra{\phi_x}{\phi_x}-I_x$. Since this is the reflection in
$\ket{x} \otimes \mathbb{C}^X $ about $\ket{\phi_x}$, this can be
written as
\[
\ketbra{x}{x} \otimes \left( U_x (2
\ketbra{\overline{0}}{\overline{0}}-I) U^{\dag}_x \right),
\]
and $2\ketbra{\overline{0}}{\overline{0}}-I$ can be implemented
using $O(\log{|X|})$ gates similarly to the construction of a
circuit for $2\ketbra{00}{00}-I$ in Section~\ref{sec:circuit}.

From the above argument, we see that if there is an efficient
procedure to implement the transformation
\[
\left(I\otimes
U_x\right)\ket{x}\ket{\overline{0}}=\ket{x}\displaystyle\sum_{y}\sqrt{p_{x,y}}\ket{y},
\]
then the algorithm can be implemented efficiently.

\subsection{Hitting Time in Quantum Walks}\label{sec:QHittingTime}
In order to analyze the deviation time, it suffices to take a look
at the eigenvalues and eigenvectors of one step of the walk. This
is because for a unitary operator $U=\displaystyle\sum_{j} e^{i
\theta_j} \ketbra{v_j}{v_j}$ with $\{\ket{v_j}\}$ being the
orthonormal eigenvectors of $U$, $U^{t}=\displaystyle\sum_{j} e^{i
\theta_j t} \ketbra{v_j}{v_j}$ so it has the same set of
eigenvectors.

Recall that $W_P=R_2 R_1= (2\Pi_2 -I)(2\Pi_1 -I)$, where $\Pi_i$
is an orthogonal projection to $E_i$, and $E_1$ is the space
spanned by $\ket{\phi_x}$ and similarly for $E_2$. Let
$A=\displaystyle\sum_{x} \ketbra{\phi_x}{x}$ and
$B=\displaystyle\sum_{y} \ketbra{\phi_y}{y}$. Note that the
dimension of the space in which $\ket{\phi_x}$ lies is $N^2$ and
that of $\bra{x}$ is $N$. Then we can write $W_P=R_2R_1$ as
$(2AA^{\dag}-I)(2BB^{\dag}-I)$ because
$\Pi_1=\displaystyle\sum_{x}\ketbra{\phi_x}{\phi_x}=AA^{\dag}$,
and similarly for $\Pi_2$. Note that $A$ and $B$ are
norm-preserving operations because $A^{\dag}A=I=B^{\dag}B$ and $A$
maps a vector in $\mathbb{C}^{x}$ into its subspace $E_1$ and
similarly for $B$.

Suppose a vector $v\in (E_1+E_2)^{\bot}$. Then $v$ lies in the
space orthogonal to both $E_1$ and $E_2$. Since $R_i$ reflects a
vector orthogonal to $E_i$, then $v$ is reflected by both $R_1$
and $R_2$. Then applying a walk operator $W_P=R_2R_1$ does not
change $v$. Hence $W_P \ket{v}=\ket{v}$, and the subspace spanned
by such $v$'s is an invariant subspace, an eigenspace with
eigenvalue $1$. Thus we only need to analyze the behavior of $W_P$
in $E_1+E_2$. Suppose we have $\ket{w}, \ket{v} \in
\mathbb{C}^{X}$, then we want to analyze the action of $R_2$ on
$A\ket{w}$ and the action of $R_1$ on $B\ket{v}$. (Since $A\ket{w}
\in E_1$, the action of $R_1$ on $A\ket{w}$ is identity.) Since
\[
\begin{array}{ll}
R_2A\ket{w}&=2\Pi_2 A\ket{w}-A\ket{w}\\
&=2B(B^{\dag}A)\ket{w}-A\ket{w},\\
\end{array}
\]
where the first term of the last line lies in $E_2$ and the second
term in $E_1$, and similarly,
\[
R_1B\ket{v}=2A(A^{\dag}B)\ket{v}-B\ket{v},\\
\]
we define the \emph{discriminant of $A$ and $B$} as follows.
\begin{defn}
The discriminant matrix $D$ of $A$ and $B$ is $D=A^{\dag}B$.
\end{defn}
\begin{thm}
If $D=\displaystyle\sum_{j} \delta_j \ketbra{w_j}{v_j}$ is the
singular value decomposition of $D$, then

a) $0\leq \delta_j \leq 1$.

b) The space generated by $\{Aw_j, Bv_j\}$, for all $j$ where
$(w_j,v_j)$ is a pair of singular vectors is invariant under
$W_P$. And $W_P$ restricted to this space is a composition of a
reflection about $Aw_j$ followed by a reflection about $Bv_j$.
\label{thm:discriminant}
\end{thm}

Let the angle between $Aw_j$ and $Bv_j$ be $\theta_j$, that is the
singular value corresponding to $Aw_j$ and $Bv_j$ be
$\bra{w_j}A^{\dag}B\ket{v_j}=\bra{w_j}D\ket{v_j}=\cos{\theta_j}$,
for $\theta_j \in [0, \pi/2]$. Recall from
Lemma~\ref{lemma:rotation} that a product of two reflections about
two reflectors $\ket{\phi}$ and $\ket{\psi}$ is a rotation by an
angle $2\theta$, where $\theta$ is an angle between the vectors
$\ket{\phi}$ and $\ket{\psi}$.  Similarly then $W_P$ is a rotation
by $2\theta_j$ in this subspace.

\begin{proof}[Theorem \ref{thm:discriminant}-a]
Singular values are taken to be real and non-negative by
convention, so $\delta_j \geq 0$. Since $A$, $B$ are
norm-preserving $\| Aw\|=\| w \|$ and $\| Bv\|=\| v \|$. Using
these facts,
\[
\begin{array}{ll}
\delta_j &\leq
\max_{\{\ket{\alpha},\ket{\beta}\}}|\bra{\beta}D\ket{\alpha}|\\
&=\max_{\{\ket{\alpha},\ket{\beta}\}}|\bra{\beta}A^{\dagger}B\ket{\alpha}|\\
&\leq \max_{\{\ket{\alpha},\ket{\beta}\}}\| A\ket{\beta}\|
\cdot\| B \ket{\alpha}\|\\
&\leq 1 .\\
\end{array}
\]
Hence $\delta_j \leq 1$. 
\end{proof}

\begin{proof}[Theorem \ref{thm:discriminant}-b]
As we mentioned before, we only need to consider the action of
$W_P$ on $Aw_j$ and $Bv_j$ and that $Aw_j$ is invariant by $R_1$
and $Bv_j$ is invariant by $R_2$.
\[
\begin{array}{ll}
W_P\ket{Aw_j} &=(2\Pi_2-I)A\ket{w_j}\\
&=(2(BB^{\dag})-I)A\ket{w_j}\\
&=2B(B^{\dag}A\ket{w_j})-A\ket{w_j}\\
&=2BD^{\dag}\ket{w_j}-A\ket{w_j}\\
&=2 \delta_j B \ket{v_j}-A\ket{w_j}\\
&=(\cos{\theta_j} B \ket{v_j})-(A\ket{w_j}-\cos{\theta_j} B
\ket{v_j}).\\
\end{array}
\]

The first term in the last line is the component of $A\ket{w_j}$
that is along $B\ket{v_j}$ and the second term is the component of
$A\ket{w_j}$ that is orthogonal to $B\ket{v_j}$. So, on
$A\ket{w_j}$, $R_1$ is a reflection about $A\ket{w_j}$ (identity
in this case), and $R_2$ is a reflection about $B\ket{v_j}$
(because of the orthogonal component) in the subspace. Similarly,
on $B\ket{v_j}$, $R_1$ is a reflection about $A\ket{w_j}$ (because
of the orthogonal component), and $R_2$ is a reflection about
$B\ket{v_j}$ (identity) in the subspace.
\end{proof}

Using Theorem~\ref{thm:discriminant}, we are ready to estimate the
deviation time with the initial state,
 \[
 \begin{array}{ll}
 \ket{\phi_{01}}&=\frac{1}{\sqrt{N}}\displaystyle\sum_{x \in
 X-M}\ket{x}\displaystyle\sum_{y\in X}\sqrt{p_{x,y}}\ket{y}\\
 &=\frac{1}{\sqrt{N}}\displaystyle\sum_{x \in
 X-M}\ket{\phi_x},\\
 \end{array}
 \]
because $\ket{\phi_{01}}$ is the state that remains after the
measurement in Step 3 of Algorithm~\ref{alg:S-Walk}.
We would like to bound $T$ such that
\[
\frac{1}{T+1}
\displaystyle\sum_{t=0}^{T}\|W_{\tilde{P}}^{t}\ket{\phi_{01}}-\ket{\phi_{01}}\|^2
\geq c(1-\epsilon)
\]
for some small positive constant $c$ and $\epsilon$ being the
fraction of marked elements. This is because by Szegedy's
definition in Definition~\ref{def:deviationTime}, the hitting time
is the time it takes for the state to be significantly different
from the initial state, greatly skewed towards the marked state.
That is, we want the $L_2$ norm of the difference between the
final and the initial state to be at least as large as the
fraction of unmarked elements. This ensures that when we measure
the final state, we detect a large deviation from the initial
state. The term in the summation is $2(1-\epsilon-\bra{\phi_{01}}
W_{\tilde{P}}^{t} \ket{\phi_{01}})$ because $
\ket{\phi_{01}}=\displaystyle\sum_{x \in
X-M}\frac{1}{\sqrt{N}}\ket{\phi_x}$ and so $\|
\ket{\phi_{01}}\|^2=1-\epsilon$, so we need to upper bound
$\frac{1}{T+1} \displaystyle\sum_{t=0}^{T}\bra{\phi_{01}}
W_{\tilde{P}}^{t} \ket{\phi_{01}}$.

Now
\[
\begin{array}{ll}
D(x,y)&=\bra{x}D\ket{y}\\
&=\bra{x}A^{\dag}B  \ket{y}\\
&=\bra{x}\left(
\displaystyle\sum_{z}\ketbra{z}{\phi_z}\right)\left(
\displaystyle\sum_{u}\ket{\phi_u}\bra{u}\right)\ket{y}\\
&=\braket{\phi_x}{\phi_y}\\
&=\bra{x}\left(\displaystyle\sum_{y}\sqrt{p_{x,y}}\bra{y}\right)\left(\displaystyle\sum_{x}\sqrt{p_{y,x}}\ket{x}\right)\ket{y}\\
&=\sqrt{p_{x,y}}\sqrt{p_{y,x}},\\
\end{array}
\]
then for $D_{\tilde{P}}$, the $(x,y)$ entry is
$\sqrt{\tilde{p}_{x,y}}\sqrt{\tilde{p}_{y,x}}$. Since if exactly
one of $x$ or $y$ is marked, $\sqrt{\tilde{p}_{x,y}}$ or
$\sqrt{\tilde{p}_{y,x}}$ is $0$, the entry of $D_{\tilde{P}}$ is
zero if exactly one of $x$ or $y$ is marked. Also since $P_M$, and
$I$ are symmetric, for $(x,y)$ both being unmarked or marked, we
have $P_M$ or $I$ respectively as diagonal blocks in
$D_{\tilde{P}}$. So
\[
D_{\tilde{P}}=\left( \begin{array}{cc}
                      P_M & 0 \\
                      0   & I \\
                      \end{array}
                \right).
\]

Now we are ready to use Theorem~\ref{thm:discriminant}. Let the
normalized eigenvectors of $P_M$ be $\{v'_{k}\}_k$ with
eigenvalues $\lambda_k$, and denote $v_k$ for $v'_{k}$ padded with
$0$ to make an eigenvector of $D_{\tilde{P}}$. Since $P_M$ is
symmetric, all the eigenvectors are orthogonal. The rest of the
eigenvectors of $D_{\tilde{P}}$ are $\{\ket{x}\}_x$ for $x \in M$.
These vectors are also orthogonal to each other, and all $n$
eigenvectors of $D_{\tilde{P}}$ are orthogonal to each other as
well. So its eigenvectors are the singular vectors and the
absolute value of the eigenvalues give the singular values because
some eigenvalues may be negative. Then from
Theorem~\ref{thm:discriminant}, the invariant subspaces of
$W_{\tilde{P}}$ are the subspaces $F_k$ spanned by the pairs
$(Av_k, Bv_k)$ with singular value $|\lambda_k |$ for all $k$ and
the subspaces $F_x$ spanned by the pairs $(A\ket{x}, B\ket{x})$
with singular values $1$ for all $x \in M$. Since the product of
two reflections is a rotation as we have seen before, the action
of $W_{\tilde{P}}$ is a rotation of the subspace $F_k$ by
$2\theta_k$, where $\theta_k$ is the angle between $Av_k$ and
$Bv_k$:
\[
\bra{v_k}A^{\dagger}B\ket{v_k}=\cos{\theta_k},
\]
This is also equal to the singular value of $A^{\dagger}B=D$
corresponding to $v_k$, $\theta_k=\cos^{-1}{|\lambda_k|}$. Also
$\theta_k \in (0, \pi/2]$ cannot be zero because
$\cos{\theta_k}=|\lambda_k|<1$  and from
Theorem~\ref{lm:delta-epsilon},
$\|P_M\|=\lambda_1=1-\frac{\delta\epsilon}{2}<1$ assuming that
$\epsilon\neq 0$. So $W_{\tilde{P}}^{t}$ rotates the subspaces by
$2\theta_k t$.

\textbf{Observation}\\
a)
\[
\ket{\phi_{01}}\in span\{Av_k\}_k
\]
because
\[
\begin{array}{ll}
\ket{\phi_{01}} &=\frac{1}{\sqrt{N}}\displaystyle\sum_{x \in
 X-M}\ket{\phi_x}\\
 &=\frac{1}{\sqrt{N}}\displaystyle\sum_{x \in
 X-M}\left(\displaystyle\sum_{z}\ketbra{\phi_z}{z}\right)\ket{x}\\
 &=\frac{1}{\sqrt{N}}\displaystyle\sum_{x \in
 X-M}A\ket{x}\\
 \end{array}
 \]
 and $\ket{x}\in span_{k}\{v_k\}$ for $n \in X-M$. So $\ket{\phi_{01}}$ is spanned
 by $Av_k$ for all $k$.

 b)
 \[
\| \ket{\phi_{01}}\|^2=1-\epsilon
\]
because
\[
\begin{array}{ll}
\|\ket{\phi_{01}}\|^2 &=\frac{1}{N}\displaystyle\sum_{x \in
 X-M}\braket{\phi_x}{\phi_x}\\
 &=\frac{|X-M|}{N}\\
 &=1-\epsilon .\\
 \end{array}
 \]

 So we normalize the initial state and also write this as the
 linear combination of the spanning set,
 \begin{equation}\label{eq:phi_01}
 \frac{\ket{\phi_{01}}}{\sqrt{1-\epsilon}}=\displaystyle\sum_{k}\nu_k A
 \ket{v_k}.
 \end{equation}

 Note that the square of the amplitudes sum up to $1$ so
 $\displaystyle\sum_{k}\nu_{k}^{2}=1$. Let $\ket{z_k}=A \ket{v_k}$. Then
 $\ket{z_k}$ are orthonormal to each other because $\ket{v_k}$
 are orthonormal to each other. $A$ preserves inner products:
 \[
 \begin{array}{ll}
 \braket{z_k}{z_{k'}} &= A^{\dagger}A \braket{v_k}{v_{k'}}\\
                      &= \bra{v_k} A^{\dagger}A \ket{v_{k'}}\\
                      &=\braket{v_k}{v_{k'}}\\
                      &= 0.\\
 \end{array}
 \]
 \begin{claim}
 If $T \geq 100 \displaystyle\sum_{k}\frac{\nu_{k}^{2}}{\theta_k}$,
\[
\begin{array}{ll}
\frac{1}{T+1}
\displaystyle\sum_{t=0}^{T}\|W_{\tilde{P}}^{t}\ket{\phi_{01}}-\ket{\phi_{01}}\|^2
&=2\left( (1-\epsilon) -
\frac{1}{T+1}\displaystyle\sum_{t=0}^{T}\bra{\phi_{01}}
W_{\tilde{P}}^{t} \ket{\phi_{01}}\right)\\
&\geq c(1-\epsilon),
\end{array}
\]
for some constant $ 2 \geq c > 0$.
 \end{claim}
This means
\[
\frac{1}{T+1}\displaystyle\sum_{t=0}^{T}\bra{\phi_{01}}
W_{\tilde{P}}^{t}
\ket{\phi_{01}} <\left( 1- \frac{c}{2} \right) \left( 1-
\epsilon\right).
\]
\begin{proof}
From Equation~\ref{eq:phi_01}
\[
\begin{array}{ll}
\bra{\phi_{01}} W_{\tilde{P}}^{t} \ket{\phi_{01}}
&=\frac{1}{1-\epsilon}\displaystyle\sum_{k_1,k_2}\nu_{k_1}\nu_{k_2}\bra{z_{k_1}}W_{\tilde{P}}^{t}\ket{z_{k_2}}\\
&=\frac{1}{1-\epsilon}\displaystyle\sum_{k}\nu_{k}^{2}\bra{z_{k}}W_{\tilde{P}}^{t}\ket{z_{k}}\\
&=\frac{1}{1-\epsilon}\displaystyle\sum_{k}\nu_{k}^{2}\cos{(2\theta_k t)},\\
\end{array}
\]
because $\ket{z_{k}}$'s are orthonormal to each other and belong
to orthogonal eigenspaces of $W_{\tilde{P}}$. The last line is
obtained from the fact that the angle between $\ket{z_{k}}$ and
$W_{\tilde{P}}^{t}\ket{z_{k}}$ is $2\theta_k t$ since one step of
$W_{\tilde{P}}$ rotates the subspace by $2\theta_k$, and so
$\bra{z_{k}}W_{\tilde{P}}^{t}\ket{z_{k}}=\cos{(2\theta_k t)}$.

Now we use three mathematical identities to bound
$\displaystyle\sum_{t=0}^{T}\cos{(2\theta_k t)}$. First,
\[
\cos{(\omega t)}=\frac{e^{i \omega t}+e^{-i \omega t}}{2}.
\]

So the sum of cosines is a sum of two geometric series. Using the
formula for the sum of geometric series we have,
\[
\begin{array}{ll}
\frac{1}{T+1}\displaystyle\sum_{t=0}^{T}\bra{\phi_{01}}
W_{\tilde{P}}^{t} \ket{\phi_{01}}&\leq
(1-\epsilon)\displaystyle\sum_{k}\nu_{k}^{2}\frac{1}{T+1}
\frac{\cos{(2\theta_kT)}- \cos{(2\theta_k(T+1))}+ 1-
\cos{(2\theta_k)}}{2(1-\cos{(2\theta_k}))}.\\
\end{array}
\]

Next, we use
\[
|\cos{\alpha}-\cos{\beta}| \leq | \alpha - \beta|
\]
to bound the numerator and use
\[
\cos{\alpha} \leq 1- \frac{\alpha^2}{8}\text{   for $\alpha \in
[-3.79, 3.79]$},
\]
to bound the denominator. Note that here, $\alpha=2\theta_k$ and
$\theta_k \in (0, \frac{\pi}{2}]$ as mentioned before, so the
third inequality can be applied.

Using these, we get
\[
\begin{array}{ll}
\frac{1}{T+1}\displaystyle\sum_{t=0}^{T}\bra{\phi_{01}}
W_{\tilde{P}}^{t}
\ket{\phi_{01}}&\leq (1-\epsilon)\frac{1}{T+1}\displaystyle\sum_{k}\nu_{k}^{2}\frac{2\theta_k + 2\theta_k}{2((2\theta_k)^2)/8}\\
&=(1-\epsilon)\frac{4}{T+1}\displaystyle\sum_{k}\frac{\nu_{k}^{2}}{\theta_k}\\
&\leq (1-\epsilon)\frac{4}{100}.\\
\end{array}
\]

The last line comes from the fact that we have chosen $T \geq 100
\displaystyle\sum_{k}\frac{v_{k}^{2}}{\theta_k}$. So the claim
holds for $\left(1-\frac{c}{2}\right)=\frac{1}{25}$ or
$c=\frac{48}{25}$.
\end{proof}

We can relate the hitting time of the walk with the eigenvalues of
$P_M$.
\begin{cor}
$c$-deviation on average time for $W_{\tilde{P}}$ with respect to
$\frac{\ket{\phi_{01}}}{\sqrt{1-\epsilon}}$ is
$O\left(\frac{1}{\sqrt{1-\|P_M\|}}\right)$. \label{cor:sqrt}
\end{cor}
\begin{proof}
We know that $T \geq 100
\displaystyle\sum_{k}\frac{\nu_{k}^{2}}{\theta_k}$ and that
$\cos{\theta_k}=|\lambda_k|$. Then
\[
\begin{array}{ll}
\theta_k \geq \sin{\theta_k} &= \sqrt{1-\cos^2{\theta_k}}\\
                             &=\sqrt{1-\lambda_{k}^{2}}\\
                             &\geq \sqrt{1-\lambda_k},\\
\end{array}
\]
because $\lambda_k \leq 1$.

So,
\[
T \leq 100
\displaystyle\sum_{k}\frac{\nu_{k}^{2}}{\sqrt{1-\lambda_{k}^{2}}}
\leq \frac{100}{\sqrt{1-\|P_M\|}}\displaystyle\sum_k \nu_{k}^{2}
\]
because $\displaystyle\sum_k \nu_{k}^{2}=1$ and any eigenvalue in
$P_M$ is at most the largest eigenvalue of $P_M$ which is
$\|P_M\|$. This means that the hitting time $T\in
O\left(\frac{1}{\sqrt{1-\|P_M\|}}\right)$.
\end{proof}

Recall that the classical hitting time for a symmetric transition
matrix is $O\left(\frac{1}{1-\|P_M\|}\right)$, so using Szegedy's
walk we have quadratic speedup.
\begin{theorem}[\cite{Sz04b}]\label{thm:Mario}
For the quantum walk based on a transition matrix $P$ with
eigenvalue gap of $\delta$, the fraction of marked elements
$|M|/|X|$ at least $\epsilon$, in time $O\left(1/\sqrt{\delta
\epsilon}\right)$, Algorithm~\ref{alg:S-Walk} detects a marked
element with probability at least $\frac{1}{1000}$ if it exists,
in $O(1/\sqrt{\delta \epsilon})$ application of $W_{\tilde{P}}$.
\end{theorem}
\begin{proof}
If a marked element exists, either a marked element is detected in
Step 2 with probability $\epsilon$, or a deviation is detected in
Step 11 with probability $\geq \frac{1}{4(T+1)}
\displaystyle\sum_{t=0}^{T}\|W_{\tilde{P}}^{t}\ket{\phi_{01}}-\ket{\phi_{01}}\|^2
\geq \frac{12}{25}(1-\epsilon)$. Then the net probability of
success is $\epsilon + \frac{12}{25}(1-\epsilon) \geq
\frac{12}{25}+\frac{13 \epsilon}{25}$. Here,
$T=O(\frac{1}{\sqrt{1-\|P_M\|}})$ from Corrollary~\ref{cor:sqrt}
and Lemma~\ref{lm:delta-epsilon}.
\end{proof}

The consequence of Theorem~\ref{thm:Mario} is that it suffices to
analyze the classical version of the walk in order to bound the
quantum hitting time. Suppose we have three different costs, time
or query, associated with a classical walk based algorithm. A
setup cost, $s(r)$, an update cost $u(r)$ and a checking cost
$c(r)$. A setup cost is the cost required to set up the initial
$r$-subset, an update cost is the cost to maintain the data
pertaining to the $r$-subset during the walk, and a checking cost
is the cost needed to see if we have a marked subset. Then the
total \emph{quantum} complexity of this algorithm is
\begin{equation}\label{equ:s-walk} s(r)+\frac{1}{\sqrt{\delta
\epsilon}}\left(u(r)+c(r)\right).
\end{equation}

Throughout the rest of the essay, we will describe the classical
versions of the algorithms to obtain quantum upper bound.

As an application to element distinctness, using
Theorem~\ref{thm:Mario}, we get $O(n^{2/3})$ bound. If we use the
classical walk, however, we get a query complexity of
$O(n^{4/3})$, which is worse than the straightforward algorithm
that gives $O(n)$. Because in quantum case, we have a smaller
hitting time, a walk based approach performs better.

\section{Quantum Walk of Ambainis} \label{sec:A-walk}
There is another quantum walk algorithm proposed by
Ambainis~\cite{Amb04} to solve Element Distinctness, which came
prior to~\cite{Sz04a, Sz04b}. This is generalized in~\cite{MSS05,
CE03} to solve any $k$-collision problem and is called
\emph{Generic Algorithm}.
\begin{defn}[$k$-collision]~\cite{CE03}
Given a function $f$ on a set $S$ as an oracle and a $k$-ary
relation $C \subseteq S^k$, find a $k$-tuple of distinct elements
$(a_1, a_2, \ldots, a_k) \in S^k$ such that $(f(a_1), f(a_2),
\ldots, f(a_k)) \in C$ if it exists. Otherwise, reject.
\end{defn}

In the circuit for the generic algorithm, there are three main
registers, a set register, a data register and a coin register.
The set register holds a subset $I$ of the set $S$, of size $r$ or
$r+1$. The data register holds the data corresponding to the set
in the set register. The coin register holds an element of $S-I$.
In element distinctness, for example, the set register contains
indices of elements $i$ in $r$-subset, the data register contains
the actual value $x_i$ for each element in the set register, and
the coin register contains the indices $j$'s that are not in the
set register.

The walk starts with a uniform superposition of $r$-subset in the
set register and sets up the corresponding data register as in
Szegedy-walk. Unlike Szegedy-walk, this algorithm also sets up a
coin register $C$. At each step of the walk, if the subset is
marked, \ie\ contains a $k$-tuple in $C$, then it flips the phase
by applying a phase flip operator similar to the one in
Section~\ref{sec:circuit}. Then it enters quantum walks to
\emph{flip the coin}. It diffuses the coin register over indices
in $S-I$ by applying a Grover diffusion operator similar to the
one in Section~\ref{sec:G} and adds the element from the coin
register to the set register. Now the size of the set register is
augmented to $r+1$. Then it diffuses the set register over $I$,
and removes one element from the set register. Note that during
this diffusion step, the data register is updated correspondingly.
This process is repeated for some time before
checking the subset for a marked state. 
When the size of $r$-subset is $1$, this is analogous to what
Grover's algorithm does.
%
%

Similarly to Equation~\ref{equ:s-walk}, we can write the
expression for the total cost of Ambainis-walk using a setup cost,
an update cost and a checking cost from the classical walk,
\[\label{equ:a-walk}
s(r)+\left(\frac{n}{r}\right)^{k/2}(c(r)+\sqrt{r}u(r)).
\]

One of the differences between Ambainis-Walk and Szegedy-Walk is
that in the former, checking takes after $\sqrt{r}$ steps of the
quantum walk, whereas in the latter, checking takes after
\emph{every} step of the walk. Also, in the former, the walk is
over a graph, in which the vertices are a subset of size $r$ or
$r+1$ and they are connected iff the size of the vertex differ by
$1$ and the symmetric difference is two, whereas in the latter,
the walk is over a bipartite graph, and each side of the vertices
are subsets of size $r$, and they are connected iff the symmetric
difference is two. We shall see later how these differences affect
the performance of an algorithm for different problems.

\section{Triangle Finding Problem} \label{sec:triangle}
Suppose we are given an oracle for the adjacency matrix of a
graph. It takes two vertices in a graph $(i,j)$ as inputs and
outputs $1$ if the vertices are connected by an edge and $0$
otherwise. We are promised that there is exactly one clique of
size three, called \emph{triangle}, or none at all. Our goal is to
test which case holds for an undirected graph $\G$ with as few
queries to the oracle as possible.

For $\G$ with $n$ vertices, classical lower bound is
$\Omega(n^{4/3}\log^{1/3}n)$~\cite{CK01}. Quantumly, the lower
bound is $\Omega(n)$~\cite{MSS05}, by the following argument.
Suppose there is a graph $\G_1$, that is formed by adding an extra
edge to one pair of the $n$ leaves in a star graph, $\G_2$. Then
there are $n^2$ possible triangles in $\G_1$. We are given an
oracle for the edges in $\G_1$; it answers ``yes'' in input
$(i,j)$ if it is part of the graph. The goal is to find an edge in
$\G_1$ that is part of $\G_1 - \G_2$. Using a lower bound for
unordered search over $n^2$ edges this takes $\Omega (n)$
quantumly as we prove later in Section~\ref{sec:Glb}. Now such an
edge forms a triangle in $\G_1$. So if we are given an algorithm
for triangle finding, we could also find an edge in $\G_1 - \G_2$.
Hence the quantum lower bound for triangle finding problem is
$\Omega(n)$.

A straightforward quantum upper bound is $O(n^{1.5})$ by running
Grover search on $n^3$ triplets of vertices, querying three times
at each iteration. Here we present an algorithm of Magniez,
Santha, and Szegedy~\cite{MSS05} that uses Ambainis-based quantum
walk and queries the oracle $O(n^{1.3})$ times. We also present an
algorithm that uses Szegedy-based quantum walk to compare its
performance with Ambainis-based quantum walk algorithm. We will
also point out why there is a difference in performance between
the algorithms that use these two different
quantum walks.

\subsection{ $O(n^{1.3})$ Algorithm Using Ambainis
Walk}\label{sec:fred-triangle} Recall from
Section~\ref{sec:A-walk} that the query complexity for solving
$k$-Collision for a set of $n$ elements by performing a quantum
walk on $r$-subsets is,
\[
s(r)+\left(\frac{n}{r}\right)^{k/2}(c(r)+\sqrt{r}u(r)).
\]

The approach in~\cite{MSS05} consists of an outer algorithm $A_o$
and a subroutine $A_s$. The input of $A_o$ is a set $V_o$ of $n$
vertices. 
The output of $A_o$ is a pair of vertices in $V_o$ that is part of
a triangle if there is one, ``reject'' otherwise. 
The input for $A_s$ is a set of $r$ vertices, $V_s$ and their
adjacency matrix as well as a vertex $v$ that is not necessarily
in $V_s$.
The output of $A_s$ is an edge called \emph{Golden Edge} in the
adjacency matrix for vertices in $V_s$ that together with $v$
forms a triangle. Then in order to find a triangle edge in the
subset in $A_o$, we only need to feed each of the $n$ vertices in
$V_o$ and an adjacency matrix for a subgraph induced by an
$r$-subset into $A_s$. A further modification is that using
Grover's search algorithm, we search for a vertex that forms a
golden edge by repeating this algorithm $\sqrt{n}$ times instead
of $n$.

Next, we analyze the query cost of $A_s$ and then $A_o$. Remember
that in $A_s$, we are given the adjacency matrix of a set of
vertices $V_s$ of size $r$. We perform a walk on $s$-subsets of
$[n]$ to find a golden edge in $V_s$.  We create a subset of size
$s$ out of $r$ vertices, and query if each of $s$ vertices is
connected to the given vertex $v$, because $v$ might come from
outside this set $V_s$. This setup cost is then $O(s)$. At each
step of the walk, we get a new vertex from $V_s$ into the subset
of size $s$, but in order to check if there is a golden edge in
the subset, we only need to query if the new vertex is connected
to $v$. So the update cost is $1$ and the checking cost is $0$.
The parameter $k$ for this instance of $k$-collision is $2$,
because we are looking for two vertices that form a triangle with
$v$, giving the total query cost of the order of
\[
s+\frac{r}{s}(\sqrt{s}).
\]

This is minimized when $s=r^{2/3}$ with $O(r^{2/3})$ query cost.

The outer algorithm $A_o$ performs a walk on $r$-subsets of
vertices of $V_o$. The data are the adjacency matrix of the
subgraphs induced by the $r$-subset. Initially we need to query
$r^2$ times to set up an adjacency matrix of the subset. At each
step of the walk, we insert a new vertex in the subset and remove
one from it. We update the adjacency matrix for the new vertex,
which costs $r$ queries. For detecting a golden edge, we invoke
$\sqrt{n}$ times the subroutine $A_s$ that costs $r^{2/3}$. Hence
the checking cost is $\sqrt{n}r^{2/3}$. The parameter $k=2$
because we are looking for two vertices that are part of a
triangle. Hence the total cost is,
\[
r^2+\left(\frac{n}{r}\right)\left( \sqrt{n}r^{2/3} + \sqrt{r}r
\right).
\]

This is minimized when $r=n^{3/5}$ giving $O(n^{1.3})$ query
complexity.

\subsection{Szegedy Walk Does Not Perform Better}\label{subsec:K-triangle}
Does using Szegedy-Walk give any advantage in query complexity for
this problem? Suppose the goal of the outer algorithm $A_o$ and
the subroutine $A_s$ are the same as in~\cite{MSS05}. Then for
$A_s$, the setup cost, update cost and the checking cost do not
change. Using $s$ as the size of the subset and $r$ as the number
of vertices in $A_s$, $\delta$ is $1/s$ from
Theorem~\ref{thm:Knu}, and $\epsilon$ is the probability that we
have two vertices that form a golden edge with $v$, so $\epsilon=
\frac{{{r-2} \choose {s-2}}}{{r \choose s}}\approx
\frac{s^2}{r^2}$, for $s\in o(r)$. Then the total cost is,
\[
s+\frac{r}{\sqrt{s}}(1),
\]
minimizing this gives $r^{2/3}$ when $s=r^{2/3}$, which is exactly
the same as in~\cite{MSS05} described in
Section~\ref{sec:fred-triangle}.

For $A_o$, the setup cost, update cost and the checking cost are
as same as in~\cite{MSS05}. Using $r$ as the size of the subset,
$\delta$ is $1/r$ from Theorem~\ref{thm:Knu}, and $\epsilon$ is
the probability that we have two vertices that form a golden edge
in the $r$-subset $V_s$. So $\epsilon= \frac{{{n-2} \choose
{r-2}}}{{n \choose r}}\approx \frac{r^2}{n^2}$, for $r\in o(n)$.
Then the total cost is of the order of,
\[
r^2+\frac{n}{\sqrt{r}}(r+\sqrt{n}r^{2/3}).
\]

However, this gives $O(n^{1.5})$ query complexity for $r=O(1)$,
the same as the straightforward application of Grover's search and
worse than in~\cite{MSS05}.

It turns out for the same setup, update and checking cost, we can
easily see which algorithm will perform better~\cite{Mag05}.
Compare the formula for $k$-collision using Ambainis-Walk
\[
\begin{array}{ll}
s(r)+\left(\frac{n}{r}\right)^{k/2}(c(r)+\sqrt{r}u(r))\\
=s(r) + \frac{n^{k/2}}{r^{k/2}}c(r) +
\frac{n^{k-2}}{r^{(k-1)/2}}u(r),
\end{array}
\]
with the one for Szegedy-Walk, where $\delta=\frac{1}{r}$ and
$\epsilon=\frac{{{n-k} \choose {r-k}}}{{n \choose r}}\approx
\frac{r^k}{n^k}$,
\[
\begin{array}{ll}
 s(r)+ \frac{n^{k/2}}{r^{(k-1)/2}}(c(r)+u(r))\\
=s(r) + \frac{n^{k/2}}{r^{(k-1)/2}}c(r)
+\frac{n^{k/2}}{r^{(k-1)/2}}u(r).
 \end{array}
\]

From these we see that Ambainis-Walk always performs better
because the second term is always better than Szegedy-Walk, while
other terms are the same. This allows us to have a higher query
cost for checking, giving an improvement over the straightforward
$O(n^{1.3})$ upper bound for triangle finding algorithm.

There are other algorithms that use Szegedy-Walk, such as an
algorithm that performs a walk based on edges. However, so far all
these algorithms give the same $O(n^{1.5})$ query upper bound.

\section{Adversary Method for Query Lower Bounds}
In this section, we describe one of the popular methods to derive
lower bounds for quantum query complexity. Later in this essay we
apply this technique to derive lower bounds for the problems we
are studying.
\subsection{Quantum Adversary Theorem} \label{sec:lowerbound}
Suppose an oracle takes an input $i$ and produces $x_i$ to form a
string $x=(x_1, x_2, \ldots, x_N)^N \in\{0,1\}$. Furthermore,
suppose there is a boolean function that takes the string $x$ as
an input and produces an output $f(x)$. For example, in unordered
search, the oracle takes an index $i$ and outputs $x_i$. The
boolean function $f(x)$ is the logical OR of all $x_i$:
$f(x)=\vee_{i}x_i$. We want to lower bound the number of queries
needed to decide $f(x)$.
\begin{thm}~\cite{Amb00}\label{thm:lb-weak}
Let $A\subseteq\{0,1\}^N$ be a set such that every string in the
set maps to $0$ under $f$, and let $B\subseteq\{0,1\}^N$ be a set
such that every string in the set maps to $1$. Suppose that
\begin{enumerate}
\item For all $x \in A$, there exists $m$ different $y\in B$ such
that $y_i \neq x_i$ for exactly one $i$.

\item For all $y \in B$, there exists $m'$ different $x\in A$ such
that $x_i \neq y_i$ for exactly one $i$.
\end{enumerate}
Then $\Omega(\sqrt{mm'})$ queries are required to compute $f$.
\end{thm}
\begin{proof}
Suppose we have a $t$ query bounded error algorithm for computing
$f$. In order to lower bound $t$, the number of queries needed, we
take a look at $W_t$, the sum of all the inner products at the end
of $t$-th query over all pairs of $x$ and $y$ that satisfy the
relationships stated in parts 1 and 2 of the theorem:
\begin{equation}\label{eq:lbinner}
W_t=\displaystyle\sum_{(x,y) \in R}\braket{\psi^t_x}{\psi^t_y}.
\end{equation}
The proof estimates the difference $|W_t-W_0|$ and
$|W_j-W_{j-1}|$, the difference made after each query to the
oracle in terms of $|R|$.

Let $\ket{\psi^j_x}$ be the state of the algorithm after the
$j$-th query if queries were answered according to the input
$x=(x_1, x_2, \ldots, x_N)$. We are interested in
$\braket{\psi^j_x}{\psi^j_y}$, \ie\ how much the states will
differ after $j$ queries if $x$ is taken from the set $A$ and $y$
is taken from the set $B$. For this inner product, there are two
simple things we know about,
\begin{description}
\item[Property 1] $\braket{\psi^0_x}{\psi^0_y}=1$. This is because
$\ket{\psi^0_x}=\ket{\psi^0_y}=\ket{\psi_{start}}$.

\item[Property 2] At the end of the algorithm, the inner product
must be small: After $t$ queries, $|\braket{\psi^t_x}{\psi^t_y}|
\leq c$ for a constant $c<1$.
\end{description}
\begin{proof}
The proof of Property 2 above follows from the lemma,
\begin{lemma}[\cite{AKN98}]
If $|\braket{\psi_1}{\psi_2}| \geq 1- \epsilon$, then for any
measurement $M$ and any outcome $i$, the probability of finding
$i$ when measuring $\ket{\psi_1}$ and $\ket{\psi_2}$ differs by at
most $\sqrt{2 \epsilon}$.
\end{lemma}
Suppose there is an algorithm with the probability of obtaining
correct outcome greater than or equal to $\frac{3}{4}$. The
probability of having an outcome $0$ is at least $\frac{3}{4}$ if
we have an input $x$ such that $f(x)=0$. The probability of
obtaining $0$ is less than $\frac{1}{4}$ if we have $y$ such that
$f(y)=1$. This means that if we have $x\in A$ and $y \in B$ and
measure the final state $\ket{\psi^t_x}$ and $\ket{\psi^t_y}$ then
the probability of measuring $0$ differs by at least $1/2$. So
\[
\begin{array}{l}
 \sqrt{2 \epsilon} \geq \frac{1}{2}\\
 \epsilon \geq \frac{1}{8}\\
 1-\epsilon \leq \frac{7}{8}.\\
 \end{array}
 \]

Therefore, the inner product $|\braket{\psi^t_x}{\psi^t_y}|$
differs by at most $\frac{7}{8}<1$.
\end{proof}
\begin{description}
\item[Property 3] From Property 1, we know
$W_0=\displaystyle\sum_{(x,y) \in
R}\braket{\psi^t_0}{\psi^t_0}=\displaystyle\sum_{(x,y) \in R}
1=|R|$, where $|R| \geq |A|m,|B|m' $.

\item[Property 4] From Property 2, we know that after the last,
$t$-th query, each of the inner product is at most $\frac{7}{8}$
in absolute value, so $|W_t| \leq \frac{7}{8} |R|$.
\end{description}
\begin{lemma} \label{claim:1} $|W_j-W_{j-1}|\leq
\frac{2}{\sqrt{mm'}}|R|$.
\end{lemma}

We will provide a proof of Lemma~\ref{claim:1} shortly. From
Property 3 and Property 4, we get $|W_t-W_0| \geq \frac{1}{8}|R|$,
that is queries performed during the entire algorithm decreases
the inner products in Equation~\ref{eq:lbinner} at least one
eighth the size of $R$. Since at each step of the query, quantity
$W_j$ decreases by at most $\frac{2}{\sqrt{mm'}}|R|$ from
Lemma~\ref{claim:1}, the total number of queries must be at least
\[
t \geq \frac{|W_t-W_0|}{\frac{2}{\sqrt{mm'}}|R|} \geq
\frac{\sqrt{mm'}}{16}.
\]

This proves the query lower bound of $\Omega(\sqrt{mm'})$.
\end{proof}

We are now left with the proof of Lemma~\ref{claim:1}.

\begin{proof}
Let
\[
\ket{\psi^{j-1}_x}=\displaystyle\sum_{i=1}^n
\alpha_{x,i}\ket{i}\ket{\phi_{x,i}},
\]
where $\ket{\psi^{j-1}_x}$ is the state of the algorithm before
$j$-th query on input $x$. After $j$-th query, we get
\[
\ket{\psi^{j}_x}=\displaystyle\sum_{i=1}^n
\alpha_{x,i}\ket{i}\ket{\phi'_{x,i}},
\]
where $\ket{\phi'_{x,i}}$ is obtained from applying a query
operator $Q$ to $\ket{i}\ket{\phi_{x,i}}$. 
Now suppose we have two input strings $x=(x_1, x_2, \ldots, x_N)$
and $y=(y_1, y_2, \ldots, y_N)$, where we have exactly one $i$
such that $x_i \neq y_i$. For such $i$, we can rewrite
$\ket{\psi^{j-1}_x}$ as the part that involves such $i$ and the
rest,
\[
\ket{\psi^{j-1}_x}=\alpha_{x,i}\ket{i}\ket{\phi_{x,i}} +
\ket{\psi'_x}
\]
and similarly for $\ket{\psi^{j-1}_y}$,
\[
\ket{\psi^{j-1}_y}=\alpha_{y,i}\ket{i}\ket{\phi_{y,i}} +
\ket{\psi'_y}.
\]

The inner product $\braket{\psi^{j-1}_x}{\psi^{j-1}_y}$ can also
be decomposed into two parts, the one that involves the $i$ and
the rest,
\begin{equation} \braket{\psi^{j-1}_x}{\psi^{j-1}_y}=\alpha_{y,i}^*
\alpha_{x,i} \braket{\phi_{y,i}}{\phi_{x,i}} +
\braket{\psi'_y}{\psi'_x} .\label{eq:j-1<>}
\end{equation}

Similarly, we can rewrite $\ket{\psi^{j}_x}$ and
$\ket{\psi^{j}_y}$ as
\[
\ket{\psi^{j}_x}=\alpha_{x,i}\ket{i}Q_{x_i}\ket{\phi_{x,i}} +
Q\ket{\psi'_x}
\]
and
\[
\ket{\psi^{j}_y}=\alpha_{y,i}\ket{i}Q_{y_i}\ket{\phi_{y,i}} +
Q\ket{\psi'_y}.
\]

The inner product of the final state is,
\begin{equation}
\braket{\psi^{j}_x}{\psi^{j}_y}=\alpha_{y,i}^* \alpha_{x,i}
Q_{y_i}^*Q_{x_i}\braket{\phi_{y,i}}{\phi_{x,i}} +
\braket{\psi'_y}{\psi'_x}. \label{eq:j<>}
\end{equation}

Note that the query in Equation~\ref{eq:j<>} does not change the
second term because we apply the same unitary transformation $Q$
to the second registers for both $\ket{\psi'_x}$ and
$\ket{\psi'_y}$. They contain the same data, and the unitary
transformation preserves inner products. So we only need to be
careful about how much $\braket{\phi_{y,i}}{\phi_{x,i}}$ changes.
Since the inner product of $\ket{\phi_{y,i}}$ and
$\ket{\psi_{x,i}}$ is at most $1$ and
$|\alpha^*_{y,i}\alpha_{x,i}| \leq |\alpha_{y,i}||\alpha_{x,i}|$,
\[
|\braket{\psi^{j}_x}{\psi^{j}_y}-\braket{\psi^{j-1}_x}{\psi^{j-1}_y}|
\leq 2|\alpha_{y,i}||\alpha_{x,i}|.
\]

However, we are interested in the difference above for all
$(x,y)\in R$, so
\[
\begin{array}{ll}
|W_j-W_{j-1}| & \leq 2 \displaystyle\sum_{(x,y) \in
R}|\alpha_{y,i}||\alpha_{x,i}|\\
&\leq \displaystyle\sum_{(x,y) \in R}\left(\gamma|\alpha_{x,i}|^2
+
\gamma^{-1}|\alpha_{y,i}|^2 \right).\\
\end{array}
\]

For going from the second to the third line above, we used an
inequality $2AB \leq A^2 + B^2$ with
$A=\sqrt{\gamma}|\alpha_{x,i}|$ and
$B=\sqrt{\gamma}^{-1}|\alpha_{y,i}|$.

Now we bound $\displaystyle\sum_{(x,y) \in
R}\gamma|\alpha_{x,i}|^2$ and $\displaystyle\sum_{(x,y) \in
R}\gamma^{-1}|\alpha_{y,i}|^2$ separately.
\[
\begin{array}{ll}
 \displaystyle\sum_{(x,y) \in R}\gamma|\alpha_{x,i}|^2
&=\gamma\displaystyle\sum_{x \in A}\displaystyle\sum_{y:(x,y) \in
R}
|\alpha_{x,i}|^2\\
&\leq \gamma\displaystyle\sum_{x \in A}1\\
&=\gamma|A|\\
&\leq \gamma\frac{|R|}{m}\\
\end{array}
\]

Above, we used the fact that given $x$, we have at most $N$
different $y$'s that differ by exactly one position:
\[
\begin{array}{ll}
\displaystyle\sum_{y:(x,y) \in R}|\alpha_{x,i}|^2 &
=\displaystyle\sum_{i=1}^{N}|\alpha_{x,i}|^2\displaystyle\sum_{y:(x,y)\in
R, x_i
\neq y_i}1\\
&\leq \displaystyle\sum_i|\alpha_{x,i}|^2\\
&=1.\\
\end{array}
\]
The last line comes from the fact that the squares of amplitudes
sum up to $1$. Also, since for every $x\in A$, we have at least
$m$ different $y \in B$ that differ by 1, so $|R|\geq m|A|$ and
hence $|A| \leq \frac{|R|}{m}$.

Similarly $\displaystyle\sum_{(x,y) \in
R}\gamma^{-1}|\alpha_{y,i}|^2 \leq \frac{1}{\gamma}\frac{|R|}{m'}$
and we get
\[
\begin{array}{ll}
|W_j-W_{j-1}| & \leq \displaystyle\sum_{(x,y) \in
R}\left(\gamma|\alpha_{x,i}|^2 +
\gamma^{-1}|\alpha_{y,i}|^2 \right)\\
&\leq\frac{\gamma}{m}|R| +\frac{\gamma^{-1}}{m'}|R|\\
&=\frac{m'\gamma+m\gamma^{-1}}{mm'}|R|.\\
\end{array}
\]

The above expression is minimized when
$\gamma=\sqrt{\frac{m}{m'}}$ to give
\[
|W_j-W_{j-1}| \leq \frac{2}{\sqrt{mm'}}|R|.
\]
\end{proof}

The Quantum Adversary Theorem we have proven is of the simplest
form in that the yes and no instances only differ in exactly one
position.
The stronger form of the previous theorem relaxes the number of
$i$ at which $x$ and $y$ differ to be more than one. This gives a
tighter bound for several problems of interest.
\begin{thm}~\cite{Amb00}\label{thm:lb-strong} For a boolean function $f:\{0,1\}^n \rightarrow
\{0,1\}$, let $A \subseteq f^{-1}(0)$, $B \subseteq f^{-1}(1)$ and
$R \subseteq A \times B$.
\begin{enumerate}
\item For all $x \in A$, $|\{y:(x,y)\in R\}|\geq m$.

\item For all $y \in B$, $|\{x:(x,y)\in R\}|\geq m'$.

\item Define $l_{x,i}=|\{y:(x,y) \in R,\; x_i \neq y_i\}|$,
$l_{y,i}=|\{x:(x,y) \in R, \; x_i \neq y_i\}|$ and
$l=max_{(x,y)\in R, i:x_i \neq y_i}\{l_{x, i}, l_{y,i}\}$
\end{enumerate}

Then $\Omega\left(\sqrt{\frac{mm'}{l}}\right)$ queries are
required.
\end{thm}

Unfortunately, it is proven by Szegedy~\cite{Sze03} and
independently by Zhang~\cite{Zh03} that this method cannot provide
a tight lower bound for all the problems. Informally, a
$1$-certificate is the least number of bits of the input that
determines the value of the function to be $1$. If the size of a
$1$-certificate is $C_1(f)$, and $N$ is the number of variables in
the boolean function to the oracle, then the method can only prove
up to the lower bound of $O(\sqrt{C_1(f) N})$~\cite{Zh03}. For
example, in element distinctness, $C_1(f)=2$, because we need to
know the two elements that collide. Then this quantity is
$O(\sqrt{n})$, but the tight lower bound of this problem is
$\Theta(n^{2/3})$ using polynomial method~\cite{AS04}.

The polynomial method~\cite{BBC+01} is another powerful lower
bound technique. However, this method is also proven not to be
tight by Ambainis~\cite{Amb03}. As far as we know neither the
adversary nor the polynomial method provides a tight lower bound
for all problems of interest. For some problem, the adversary
method provides a better bound than polynomial method~\cite{Amb03}
and the opposite also holds~\cite{AS04}.

\subsection{The Graph Connectivity}\label{subsec:Graph-connectivity}
As an application of Theorem~\ref{thm:lb-strong}, we take a look
at the Graph Connectivity problem~\cite{DML03}. An undirected
graph $\G$ is described by $n \choose 2 $ variables $\{G_{i,j}\}$,
where $G_{i,j}=1$ if $(i,j)$ is an edge in $\G$ and $0$ otherwise.
The oracle gives the entries of adjacency matrix $G_{i,j}$. We
want to find if $\G$ is connected by making as few queries to
$G_{i,j}$ as possible. What would be the lower bound for quantum
query complexity?
\begin{figure}[h]
\begin{center}
\includegraphics[height=3in, width=4in]{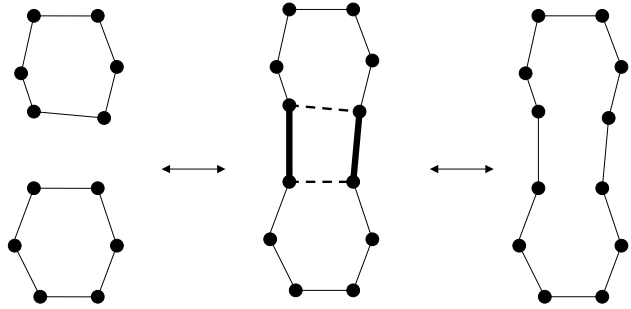}
\caption{Transformations between $G$ and $G'$}
\end{center}
\end{figure}

Let $A$ be the set of graphs on $n$ vertices that consist of two
cycles not connected one to another, each cycle of length at least
$n/3$. Let $B$ be the set of graphs that are one cycle of length
$n$. In both cases each vertex belongs to one of the cycles. We
define the relationship as $R=\{\text{$(G,G'):G'$ has exactly two
edges not in $G$}\}$. We can obtain $G'$ from $G$ by deleting one
edge from each cycle in $G$ and inserting two edges to make it a
single cycle. When connecting cycles, there are two ways, cross or
parallel. So given $G$, the number of possible $G'$ you can make
is
\[
\begin{array}{ll}
|\{G':(G, G') \in
R\}|&=(\text{length of first cycle})(\text{length of second cycle})\times 2\\
&\geq 2\frac{n}{3}\frac{n}{3}\\
&=\frac{2n^2}{9}\\
\end{array}
\]
because each cycle in $G$ is of length at least $n/3$. Hence
$m=\Omega(n^2)$.

Creating $G$ from $G'$ starts by picking one edge out of $n$ edges
in the cycle. Since each cycle in $G$ is of length at least $n/3$,
the next one must be at distance at least $n/3$ from the edge we
just picked. This leads to $n-2(n/3)=n/3$ choices for the second
edge. After that there is only one way to connect the vertices to
create the two cycles. Hence $m'=\Omega(n^2)$.

For each instance in $G$, the number of instances in $G'$ that
differs at position $(i,j)$, \ie\ $l_{G,(i,j)}$ is $O(n)$ or
$O(1)$. If $(i,j)$ is an edge in $G$, and $(i,j)$ is not an edge
in $G'$, then there are $\leq 2n/3$ graphs $G'$'s we can make by
removing $(i,j)$ and one of at most $2n/3$ edges in $G$, and we
have $l_{G,(i,j)}=O(n)$. If $(i,j)$ is not an edge in $G$, but an
edge in $G'$, then there are four ways to create $G'$ from $G$,
\eg\ by connecting $(i,j)$ and connecting a vertex to the left of
$i$ with the one left of $j$ or connecting a vertex right of $i$
to the one to the right of $j$. Similarly, $l_{G',(i,j)}=1$ if
$(i,j)$ is an edge in $G$, and at most $n$ otherwise. Overall then
we have $l \leq O(n)$, and the query complexity is
$\Omega\left(\sqrt{\frac{mm'}{l}}\right)=\Omega\left(\sqrt{\frac{n^2n^2}{n}}\right)=\Omega(n^{1.5})$.

\subsection{Lower Bound for Unstructured Search}\label{sec:Glb}
In this section, using Theorem~\ref{thm:lb-weak}, we prove a lower
bound for a search on unstructured database~\cite{Amb03}.
Unordered search is defined as follows. Given an oracle for
$x=(x_1, x_2, \ldots, x_n)\in \{0,1\}^n$, is there $i$ such that
$x_i=1$? This lower bound is useful in later sections when we
reduce from this search problem to the problems of our interest.
This lower bound was first proven in~\cite{BBBV97} using a
``hybrid argument''.
\begin{thm}~\cite{Amb03}\label{thm:lb-search}
The query complexity of a search on unstructured database of size
$n$ is $\Omega(\sqrt{n})$.
\end{thm}
\begin{proof}
Suppose we have $n$ boolean elements, $(x_1, x_2, \ldots, x_n)$.
Let $A$ be the set that contains exactly one $x_i=1$ for some
$i\in[n]$. Let $B$ be the set such that $x_j=0$ for all $j$. Then
for every $a\in A$, there are $m=1$ elements in $B$ that differ by
exactly one position. For every $b\in B$, there are $m'=n$
different elements in $A$ that differ by exactly one position.
Using Theorem~\ref{thm:lb-weak}, the number of queries needed to
search an element in unstructured database is $\Omega(\sqrt{n})$.
\end{proof}

\section{Quantum Matrix Verification
Problem}\label{sec:matrix-ver} Suppose we want to verify if $AB=C$
for $n\times n$ matrices $A$, $B$, and $C$ over some ring. The
oracle knows the entries of $A$, $B$, and $C$. What is the query
and time complexity for this problem? Classically, there is an
$O(n^2)$ time algorithm by Freidvals using random
vectors~\cite{Fre79}. Classical query lower bound for this problem
is $\Omega(n^2)$, by a reduction from unordered search; Let $A$
and $B$ be matrices having all entries being $1$: Let $C$ be a
matrix with all entries being $n$. Then $AB=C$. If we set one of
the $3n^2$ entries to be $0$ then $AB=C$ no longer holds. Hence we
are searching for one entry of $0$ out of $3n^2$ entries. The
classical lower bound for unordered search for $n^2$ elements is
$\Omega(n^2)$, hence we have an $\Omega(n^2)$ lower bound for
matrix verification.

\subsection{Upper Bound}\label{sec:Matrix-ub}
An $O(n^{5/3})$ query upper bound can be obtained by using either
Ambainis-Walk or Szegedy-Walk. The idea behind this is to perform
a walk over $r$-subsets from the set of rows from $A$ and another
$r$-subsets of a set of columns from $B$, and the corresponding
entries from $C$. For an $n \times n$ matrix $M$ and an $r$-subset
$S$ of $[n]$, let $M|_S$ denote a $r \times n$ sub-matrix of $M$
corresponding to rows in $S$, $M|^S$ an $n \times r$ sub-matrix of
$M$ corresponding to columns in $S$. Initially, we query $r$ rows
of $A$, $r$ columns of $B$ and $r^2$ entries of $C$ corresponding
to all these rows and columns. So the setup cost is $O(rn)$. When
update, we swap in a new row for $A$, a new column for $B$ and
$2r$ entries of $C$, giving the update cost of $O(n)$. Checking is
done by performing $A|_S \times B|^T$ to see if it is equal to
$C|_{S}^{T}$ for subsets $S$ and $T$. Then the checking cost is
$0$. Here, we are looking for $k=2$ elements, an index for a row
in $A$ and an index for a column in $B$ that gives a wrong entry
in $C$. The total cost if we use Ambainis-Walk is of the order of
\[
rn +\left( \frac{n}{r} \right)^{2/2}(\sqrt{r}n).
\]

Since we are looking for two elements that collide,
$\epsilon\approx \frac{r^2}{n^2}$ for $r\in o(n)$ and the spectral
gap of the walk is $\frac{1}{r}$. Then the query cost if we use
Szegedy-Walk is of the order of
\[
rn + \frac{n}{\sqrt{r}}(n).
\]

Here we see that both formulae give the same result, an
$O(n^{5/3})$ query upper bound when $r=n^{2/3}$.

Buhrman and Spalek~\cite{BS04} showed another Szegedy-Walk
algorithm that uses random vectors to speed up the running time of
the algorithm, the query complexity stays the same. In the
original Szegedy-based algorithm described above, multiplying
$A|_S$ with $B|^T$ takes $O(nr^2)$ multiplications. This time can
be reduced by using Freivalds' random vector technique on
sub-matrices. At a setup stage, we multiply $A|_S$ with a vector
$u$ of length $r$ and $B|^T$ with another vector $v$ of length $r$
as well as computing $u C|^{T}_{S} v$. During the walk stage we
keep updating these three vectors. At the checking stage, the
product of $uA|_S$ and $B|^Tv$ is tested against $u C|^{T}_{S} v$.
Then the setup cost is $2rn+r^2=O(rn)$, the update cost is
$2n+4r=O(n)$ (a factor of two came from erasing and rewriting
data), and the checking cost is $O(n)$.  Note that we still need
to query the same number of entries, \ie\ $O(rn)$ entries, in the
matrices as the original algorithm, and so the query complexity
stays the same. Thus we focus on how much speed up there is in
time complexity.  The marked element is a pair $(i,j)$ of a row of
$A$ and a column of $B$ such that when matrix $A$ and $B$ are
multiplied together via random vectors, it gives the incorrect
entry of $C$ at $(i,j)$. Note that since we are using random
vectors, the fraction of marked elements and the fraction of
elements that actually contribute to the product inequality, call
them \emph{visible marked element} are different. 
It can be shown, however, that the fraction of marked elements is
close to the fraction of visible marked elements, and that we can
minimize the error probability by calling this algorithm for a
constant number of times, each time picking $u$ and $v$ randomly.
Therefore, $\epsilon \approx \frac{r^2}{n^2}$ for $r \in o(n)$.
The eigenvalue gap $\delta=\frac{1}{r}$ as before, from
Theorem~\ref{thm:Knu}. The time complexity of one run of the
algorithm is
\[
rn + \frac{n}{\sqrt{r}}(n),
\]
which is $O(n^{5/3})$ when $r=n^{2/3}$. This algorithm is invoked
for a constant number of times, hence the overall time complexity
is also $O(n^{5/3})$. Algorithm~\ref{alg:BS04} describes the
classical version of their algorithm.
\begin{algorithm}\caption{A Classical Algorithm for Testing If
$AB=C$} \label{alg:BS04}
\begin{algorithmic}[1]
 \STATE Create a random $r$-subset $S$ of
 rows of $A$ and another random $r$-subset $T$ of columns of $B$.
 \STATE Pick a random $1 \times r$ row vector $u$ and a random $r
 \times 1$ column vector $v$.
 \STATE Compute $uA|_S$, $B|^{T} v$ and $u C|_{S}^{T} v$.
   \WHILE{$t \leq T_0$} 
   {
    \STATE Swap one row of $A$ and one column of $B$ chosen u.a.r.
    \STATE Recompute $uA|_S$, $B|^{T} v$ and $u C|_{S}^{T} v$.
    \STATE Test if $uA|_S \times B|^{T} v = u C|_{S}^{T} v$.
  }
    \ENDWHILE
 \STATE Answer ``AB=C''
\end{algorithmic}
\end{algorithm}

\subsection{Lower Bound}\label{matrix-lb}
We use quantum adversary theorem to prove an $\Omega(n^{1.5})$
lower bound~\cite{Amb05}. First consider a problem to test if
$Au=v$, where $A$ is an $n \times n$ matrix, $u$ is a vector of
length $n$ with all the entries being $1$, and $v$ is a vector of
length $n$ with all the entries being $n/2$. Let a matrix $A$ be
\emph{balanced} if each of its rows contains exactly $n/2$ entries
that are $1$ and exactly $n/2$ entries that are $0$. Let
\emph{unbalanced} $A$ to be such that $n-1$ rows contain exactly
$n/2$ entries of $1$ but one row contains $n/2+1$ entries of $1$.
Then for a balanced $A$, we have $Au=v$, but for an unbalanced
$A$, we have $Au \neq v$. There are $m=n(n/2)$ ways to transform a
balanced matrix $A$ into an unbalanced matrix by choosing one of
$n(n/2)$ entries that are $0$. There are $m'=n/2+1$ ways to
transform an unbalanced $A$ into a balanced $A$ by choosing one of
$n/2+1$ entries that are $1$. 
The parameter $l=1$ since balanced $A$ and unbalanced $A$ differs
by exactly one position. Hence we obtain
$\sqrt{\frac{n(n/2)(n/2+1)}{1}}=\Omega(n^{1.5})$ query lower bound
for testing if $Au=v$. Let $B$ consist of $n$ entries of $u$ in
the columns and $C$ to consist of $n$ entries of $v$ in the
columns, then the above argument still holds, and so the lower
bound for testing if $AB=C$ is $\Omega(n^{1.5})$.

\chapter{Testing Commutativity of Matrices}\label{chap:Matrix-testing}
Suppose we have $k$ matrices of dimension $n\times n$. The entries
of the matrix are given by an oracle with the input being a
triplet $(i,j,l)$ and the output being the $(i,j)$ entry of $l$-th
matrix. We want to test if all the matrices in the set commute
with each other or not by making as few queries to the oracle as
possible. Classically, we need to query all the entries of the
matrices by the following argument. Suppose all the matrices in
the set contained all $1$ entries. Then $AB=BA$ for every pair.
However, for every pair $A,B$, if we flip one of the $kn^2$
entries, say in matrix $A$, to $0$ then $AB\neq BA$ for every
other matrix $B$. Hence we have reduced the problem of unordered
search among $kn^2$ items to testing commutativity, giving the
lower bound of $\Omega(kn^2)$. Quantumly, an unordered search of
$n$ elements takes $\Omega(\sqrt{n})$ queries from
Theorem~\ref{thm:lb-search}~\cite{Amb03}, then by reduction,
quantum query complexity
of this problem is $\Omega(\sqrt{kn^2})$. 
What would be the quantum query complexity of testing the
commutativity of $k$ matrices of size $n \times n$?

\section{Commutativity Testing for a Single
Pair}\label{sec:MTSinglePair} Suppose we only want to test a
single pair of matrices, that is to see if $AB=BA$ for two $n
\times n$ matrices $A$ and $B$. The lower bound is obtained by the
reduction from the unordered search as in at the beginning of
Section~\ref{chap:Matrix-testing} with $k=1$. So quantum query
lower bound is $\Omega(n)$.
The upper bound is obtained from a modification of matrix
verification algorithm in~\cite{BS04}. When checking, instead of
testing $uA|_S \times B|^T v =u C|^{T}_{S} v$, we test $uA|_S
\times B|^T v =u B|_{S} \times A|^{T} v$. This does not affect the
overall time or query complexity of~\cite{BS04} in
Section~\ref{sec:matrix-ver}, and hence we have $O(n^{5/3})$ upper
bound for testing $AB=BA$.
%

\section{Commutativity Testing of $k$
Matrices}\label{sec:Commutativity} Now let's take a look at the
cases where we have $k$ matrices to test the commutativity. In
presenting the quantum algorithms, we will describe the classical
versions, as from Theorem~\ref{thm:Mario}, we only need to know
the classical algorithm to bound the quantum complexity.

\subsection{Two Straightforward Algorithms}\label{sec:Commutativity-naive}
The first algorithm performs a Grover search over all $O(k^2)$
pairs of matrices, at each step running a single pair
commutativity testing algorithm that costs $O(n^{5/3})$. Recall
that the single pair commutativity testing algorithm in
Section~\ref{sec:MTSinglePair} was obtained from the modification
of the bounded error matrix verification algorithm in
Section~\ref{sec:matrix-ver}. Then we have a bounded-error oracle.
However, using the Theorem of~\cite{HMdW03} in
Section~\ref{sec:error}, we can perform a quantum search with a
bounded-error oracle with the same complexity as that with a
perfect oracle. Hence, the query complexity of this algorithm is
$O(kn^{5/3})$.
%

In the second algorithm, Algorithm~\ref{alg:2} presented in the
table below, we query fewer number of matrices by querying more
entries per matrix.
\begin{algorithm}
\caption{A Classical Version of the Second Straightforward
Algorithm} \label{alg:2}
\begin{algorithmic}[1]
\STATE Create a random subset of $r$ matrices.
 \STATE Query all the entries of the matrices in the subset.
   \WHILE{$t \leq T$}
   {
    \STATE Pick a matrix to be swapped u.a.r. from the subset and swap this with the one not in the
    subset also picked u.a.r.
     \STATE For the new matrix in the subset, query all the
     entries.
     \STATE Check if all the matrices in the subset commute or
     not.
     \IF{There is a non commutative pair in the subset}

     \PRINT \texttt{``Non commutative.''}
     \RETURN

     \ENDIF
  }
    \ENDWHILE
 \STATE Answer ``Commutative''
\end{algorithmic}
\end{algorithm}
In order to estimate the query, but not time complexity, we need
to calculate the setup cost, update and checking cost, and $T$ the
number of iterations as in Section~\ref{sec:QHittingTime}. The
setup cost is $rn^2$ by querying all the entries of $r$ matrices
in the subset. The update cost is $n^2$ because we only need to
query all the entries for the new matrix we swap into the subset.
The checking cost is $0$. $T=\frac{k}{\sqrt{r}}$ because from
Theorem~\ref{thm:Mario}, $T=\frac{1}{\sqrt{\delta \epsilon}}$ and
$\delta=\frac{1}{r}$ from Theorem~\ref{thm:Knu} and
$\epsilon=\frac{{(k-2) \choose (r-2)}}{{k \choose r}}\approx
\frac{r^2}{k^2}$ for $r\in o(n)$, because we are looking for two
matrices that does not commute. Applying these costs into
Equation~\ref{equ:s-walk},
\[
rn^2+\frac{k}{\sqrt{r}}(n^2).
\]

Optimizing this, we have $r=k^{2/3}$ and hence the query
complexity is $O(k^{2/3}n^2)$.

Notice that we could also think of this problem as element
distinctness. Suppose that each element is a matrix, then we have
a collision if two matrices do not commute. Since element
distinctness can be solved in $O(k^{2/3})$ and we need to query
each of $O(n^2)$ entries of the pair of matrices in question, this
gives $O(k^{2/3}n^2)$ query complexity.

It is interesting to realize that although we could get the query
upper bound using Szegedy-Walk, we could simply apply a Grover's
search with a single pair matrix verification algorithm for the
first algorithm, and element distinctness for the second
algorithm. It seems we have not yet taken an advantage of
Szegedy-walk.

\subsection{Walk Over Separate Rows and Columns}\label{sec:commutativity-third}
The first straightforward algorithm repeatedly performs a walk
over a set of rows of matrices. What if we walk  over the rows and
columns taken from all $k$ matrices put together?
Algorithm~\ref{alg:3} describes the classical version of the walk.
This algorithm keeps two different $r$-subsets, one for rows and
one for columns. An element of $r$-subset for rows consists of
$(i,l)$, an $i$-th row of $l$-th matrix, also denoted $M_{i,l}$.
An element of $r$-subset for columns consists of $(j,m)$, a $j$-th
column of $m$-th matrix, also denoted $M^{j,m}$. This is because
we are looking for a pair of matrices $(l,m)$ and pairs of rows
and columns $(i,j)$ that do not commute \ie\ $M_{i,l}\times
M^{j,m} \neq M_{i,m} \times M^{j,l}$, and so we need to separate
all the rows and columns in different matrices. At each step of
the walk, we pick one row and one column in the $r$-subsets and
those not in the $r$-subsets u.a.r. and then swap these and update
the data registers accordingly. At the checking step, the
algorithm checks to see if there are rows $i$ and columns $j$ from
two different matrices $A$ and $B$. If so, we check the
commutativity by multiplying the $i$-th row of $A$ with $j$-th
column of $B$, and see if it agrees with the product of $i$-th row
of $B$ with $j$-th column of $A$.
\begin{algorithm}
\caption{A Classical Walk Over Separate Rows and Columns}
\label{alg:3}
\begin{algorithmic}[1]
\STATE Create an $r$-subset of rows by randomly choosing $r$ rows
among all the rows in $k$ matrices. Similarly create another
$r$-subset of columns.
 \STATE Query all the entries of the rows and columns in the subset.
   \WHILE{$t \leq T$}
   {
    \STATE Pick a row and a column u.a.r. from the
    $r$-subsets, and another row and column not in the $r$-subsets and swap these.
     \STATE For the new row and column in the subset, query all the
     entries.
     \STATE Check if there are rows $i$ and columns $j$ from two matrices $A$ and
     $B$. If so, check if the product of row $i$ of matrix $A$
     with the column $j$ of matrix $B$ is the same as that of row
     $i$ of matrix $B$ and the column $j$ of matrix $A$.
     \IF{There is a non commutative pair in the subset}
     { \PRINT \texttt{``Non commutative.''}
       \RETURN
     }
     \ENDIF
  }
    \ENDWHILE
 \STATE Answer ``Commutative''
\end{algorithmic}
\end{algorithm}

The setup cost is $O(rn)$ because we have $r$ rows and $r$ columns
in the subsets. The update cost is $O(n)$, because we need to
query one row and one column. The checking cost is $0$. We have
two walks going on over row indices and column indices, each of a
subset of size $r$. Then each walk operator has an eigenvalue gap
of at least $\frac{1}{r}$, with $\lambda_1=1, \lambda_2 \leq
1-\frac{1}{r}$. Since the eigenvalues of a tensor product of two
matrices are the products of all the pairs of eigenvalues from the
matrices, the largest eigenvalue is still $1 \cdot 1=1$ and the
second largest eigenvalue is at most $1\cdot \frac{1}{r}=
\frac{1}{r}$. Hence the eigenvalue gap of the tensor product of
the two matrices is $\delta \geq \frac{1}{r}$. The probability of
having marked elements is the probability that we have
noncommutative rows from two noncommutative matrices in the subset
of rows times the probability that we have noncommutative columns
from two noncommutative matrices in the subset of columns. Hence
$\epsilon=\left( \frac{{{nk-2} \choose {r-2}}}{{nk \choose r}}
\right)^2 \approx \frac{r^4}{n^4k^4}$ for $r \in o(nk)$. Hence our
query complexity is
\[
rn + \frac{n^2k^2}{r^{3/2}}(n).
\]

Optimizing this gives $O(k^{4/5}n^{9/5})$ for $r=k^{4/5}n^{4/5}$
when $r = o(nk)$.

Note that when $k=n$, The first two straightforward algorithms
both give $n^{8/3}$, and Algorithm~\ref{alg:3} gives
$O(n^{13/5})$, hence Algorithm~\ref{alg:3} has a better query
complexity. However, when $k<n^{2/3}$, the first straightforward
algorithm in Section~\ref{sec:Commutativity-naive} performs the
best and when $k>n^{3/2}$, Algorithm~\ref{alg:2} performs the
best.

\subsection{Simultaneous Quantum
Walk}\label{sec:Commutativity-fourth} Recall that in the first
straightforward algorithm we repeatedly performed a walk over rows
and columns of a fixed pair of matrices but no walk was performed
over the matrices. In Algorithm~\ref{alg:2}, we performed a walk
over matrices, but no walk was performed over the rows. What if we
perform a walk over matrices and rows/columns at the same time?
This is what Algorithm~\ref{alg:4} does. The quantization of
Algorithm~\ref{alg:4} gives us another $O(k^{4/5}n^{9/5})$ upper
bound. Note that it has the same query complexity as that of
Algorithm~\ref{alg:3} from the previous section.
\begin{algorithm}
\caption{A Classical Simultaneous Walk} \label{alg:4}
\begin{algorithmic}[1]
\STATE Create an $r$-subset of matrices $S$, an $s$-subset of rows
$R$, and another $s$-subset $C$ of columns.
 \STATE Query all the entries of the rows and columns in $R$ and $C$ of the matrices in the subset $S$.
   \WHILE{$t \leq T$}
   {
    \STATE Swap one matrix in the subset $S$ with the one not in the
    subset chosen u.a.r.
    \STATE For the new matrix in the subset, query the $s$ rows and
    columns in $R$ and $C$.
    \STATE Swap one row and column in the subsets $R$ and $C$ with the ones not in the
    subsets both chosen u.a.r.
     \STATE For the new row and column in each of the matrices in the subset $S$, query all the
     entries.
     \STATE Check if all the sub matrices given by the subset commute or
     not.
     \IF{There is a non commutative pair in the subset}
     { \PRINT \texttt{``Non commutative.''}
       \RETURN
     }
     \ENDIF
  }
    \ENDWHILE
 \STATE Answer ``Commutative''
\end{algorithmic}
\end{algorithm}

In Algorithm~\ref{alg:4}, we maintain two different $s$-subsets
for rows and columns. We keep all the rows and columns from all
the matrices in the $r$-subset from the same set of row indices
and column indices as the data. So the idea behind the algorithm
is to keep updating the set of indices for matrices, rows, and
columns. At each step of the walk, we get a new matrix and query
the entries of this new matrix. Then for each matrix in the
$r$-subset, we update a row and a column. Then the setup cost is
$O(rsn)$ for querying each entry of an $s \times n$ submatrix for
each matrix in $r$-subset. The update cost is $O(rn+sn)$, $O(sn)$
for a new matrix we just swapped in, and $O(rn)$ for a new row and
a column for each matrix in $r$-subset. The checking cost is $0$
because checking is done by computing the product of submatrices
whose entries we already know. We now calculate $\delta$. Let $P$
be the operator acting on matrix indices and $Q=Q_r \otimes Q_c$
be the operator acting on row and column indices. The eigenvalue
gap for $P$ is $1/r$ and for $Q$ is $1/s$. Then $\delta=\min\{1/r,
1/s\}$. The probability of having noncommutative submatrices is
$\epsilon=\left(\frac{{{k-2} \choose {r-2}}}{{k \choose r}}\right)
\left(\frac{{{n-1} \choose {s-1}}}{{n \choose s}}\right)^2$ for
$r\in o(k)$ and $s\in o(n)$. Thus we have a total query cost of
\[
rsn+\frac{kn}{rs}\sqrt{\max\{r,s\}}(rn+sn).
\]

Since $r\in o(k)$ and $s \in o(n)$, minimizing this gives
$O(k^{4/5}n^{9/5})$ with $r=s=k^{2/5}n^{2/5}$ when $k^{2/3}\leq n
\leq k^{3/2}$, $O(kn^2)$ with $r=s=1$ otherwise.

Note that walking for multiple steps before checking mixes the
elements of subsets well without changing the eigenvalue gap. Then
can we do better if the underlying classical Markov Chain is
$P^u\otimes Q^v$, that is, perform $u$ steps of the walk $P$ over
the matrices and then $v$ steps of the walk $Q$ over the
rows/columns indices? It turns out that the increased cost of
updating diminishes any gain from having the same eigenvalue gap.
\begin{thm}
Having $M=P^u\otimes Q^v$ for positive $u$ and $v$ as an
underlying classical Markov Chain does not give any better query
complexity than having $M'=P\otimes Q$.
\end{thm}
\begin{proof}
We still have the same setup, the checking cost and $\epsilon$ as
before. So the setup cost is $O(rsn)$, the checking cost is $0$
and $\epsilon=\left(\frac{{{k-2} \choose {r-2}}}{{k \choose
r}}\right) \left(\frac{{{n-1} \choose {s-1}}}{{n \choose
s}}\right)^2$ for $r\in o(k)$ and $s \in o(n)$. The update cost
this time is $(usn+vrn)$. We need to analyze the eigenvalue gap of
$M=P^uQ^v$. From Theorem~\ref{thm:Knu}, the upper bound of the
eigenvalue gap is $1/r$, hence the second largest eigenvalue is at
least $1-1/r$. Then the largest eigenvalue of $P^u$ is still $1$
and its second largest eigenvalue is at least $(1-1/r)^u$.
Similarly, the second largest eigenvalue of $Q^v$ is at least
$(1-1/s)^v$. Then the largest eigenvalues for $P^uQ^v$ is still
$1$ and the second largest is at most $\max\{(1-1/r)^u,
(1-1/s)^v\}$. Then $\delta \geq \min\{1-(1-1/r)^u, 1-(1-1/s)^v\}$.
Then we have
\[
\begin{array}{ll}
 T &=\frac{1}{\sqrt{\delta \epsilon}}\\
   &=\frac{kn}{rs}\max\left\{\frac{1}{\sqrt{1-(1-\frac{1}{r})^u}},\frac{1}{\sqrt{1-(1-\frac{1}{s})^v}}
   \right\}. \\

 \end{array}
 \]

 Hence we have
 \[
 rsn+(usn+vrn)\frac{kn}{rs}\max\left\{\frac{1}{\sqrt{1-(1-\frac{1}{r})^u}},
 \frac{1}{\sqrt{1-(1-\frac{1}{s})^v}}\right\}.
 \]

Next, we express $r$ and $s$ in terms of $k$ and $n$ that gives
the optimal bound.

We first note that $(1-1/r)^u\approx1+u(-1/r)=1-u/r$ for
$r=\omega(1)$ by taking the first two terms of binomial expansion.
Hence $\sqrt{1-(1-\frac{1}{r})^u}\approx\sqrt{u/r}$. Then we get
the following bound for the cost,
 \[
 rsn+(usn+vrn)\frac{kn}{rs}\max\left\{\frac{\sqrt{r}}{\sqrt{u}},\frac{\sqrt{s}}{\sqrt{v}}\right\}.
 \]

 Suppose $r/u \geq s/v$, then $r \geq su/v$ and $vrn \geq usn$.
 Then we get
 \[
 rsn +vrn\frac{kn}{rs}\frac{\sqrt{r}}{\sqrt{u}}.
 \]

 Simplifying this, we get
 \[
 rsn+\frac{kn^2v\sqrt{r}}{s\sqrt{u}}.
 \]

Both the first and the second terms of the sum above is an
increasing function of $r$, so we want to set $r$ to be the
minimum. Since  $r \geq su/v$, we set $r= su/v$. The new
simplified formula is then,
 \[
\frac{s^2un}{v}+\frac{kn^2\sqrt{v}}{\sqrt{s}}.
 \]

Since the first term of the sum above is an increasing function of
$s$ but the second term is a decreasing function of $s$, we set
the first term to be equal to the second term,
\[
\frac{s^2un}{v}=\frac{kn^2\sqrt{v}}{\sqrt{s}}.
 \]

Solving this gives $s=\frac{k^{2/5}n^{2/5}v^{3/5}}{u^{2/5}}$, and
the query complexity is $O(k^{4/5}n^{9/5}v^{1/5}u^{1/5})$ for
$k^{2/3}\frac{v}{u^{2/3}}\leq n \leq k^{3/2}\frac{v^{3/2}}{u}$.
Otherwise, we get $r=s=1$ with complexity $O(kn^2v)$. Similar
arguments holds for when $r/u \leq s/v$. We see that since $u$ and
$v$ are positive, the best upper bound achieved by applying
$M=P^uQ^v$ does not give any better query bound than simply
applying $M'=PQ$.
\end{proof}

\section{Generalization of Simultaneous Quantum
Walks}\label{sec:commutativity-general} In the previous problem of
testing the commutativity of $k$ matrices in
Section~\ref{sec:Commutativity}, the marked state depended on two
parameters, a set of matrix indices and the set of row/column
indices. The best upper bound was obtained by a simultaneous walk
over these two sets of indices. Suppose now the condition of being
marked depends on $m$ parameters. Then we can obtain a better
upper bound than straightforward application of Grover's search or
that of quantum walk by having a walk in each of $m$ subsets in
parallel, at each step of the walk, updating each of the
parameters. For example, for the commutativity testing of a matrix
set, $m=2$ and so at each step, we updated a matrix set and a
row/column set. The setup, the update and the checking cost, as
well as $\epsilon$ depends on how the data are stored. However,
$\delta$ is the minimum eigenvalue gap among all the walk
operators. Hence if we have $m$ subsets of size $r_1,r_2, \ldots,
r_m$, then $\delta=\min_i\{\frac{1}{r_i}\}$. Below is an example
problem that is reduced to testing the commutativity of $k$
matrices problem by having only one element in each set.

\subsection{Example Problem}\label{sec:commutativity-general-ex}
Suppose we have $m$ sets of matrices, each containing $k$ matrices
of size $n\times n$. We are promised that within each set, the
matrices commute. Are there two or more sets, when combined, give
a noncommutative set of matrices? 

\subsection{Upper Bound}\label{sec:commutativity-general-ex-ub}
The following is an $O(m^{6/7}k^{6/7}n^{13/7})$ algorithm by a
simultaneous quantum walk over the sets, matrices and
rows/columns.
\begin{algorithm}
\caption{A Classical Algorithm for Solving Collisions with Three
Parameters}\label{alg:5}
\begin{algorithmic}[1]
\STATE Create a $t$-subset $S$ of sets, $r$-subset $M$ of matrices
and $s$-subsets $R$ and $C$ of rows/columns.
 \STATE Query all the entries of the rows and columns in $R$ and $C$ of matrices in $M$
 that are in sets $S$.
   \WHILE{$t \leq T$}
   {
    \STATE Swap one set in $S$ with one not in $S$ by choosing
    the elements u.a.r.
    \STATE  Query $s$ rows and columns in $R$ and $C$ for all the
     $r$ matrices in $M$ in the new $t$-subset.
     \STATE Swap one matrix in $M$ with the one not in $M$ both
     chosen u.a.r.
     \STATE Query $s$ rows and columns in $R$ and $C$ for the new matrix in each
     of $t$ sets in $S$.
     \STATE Swap one row and column in $R$ and $C$ with the ones not in $R$ and $C$ both
     chosen u.a.r.
     \STATE Query a row and a column for the new row and column in each of
     $r$ matrices in $M$ in $t$ sets in $S$.
     \STATE Check if all the matrices in the subset commutes or
     not.
     \IF{There is a non commutative pair in the subset}
     { \PRINT \texttt{``Non commutative.''}
       \RETURN
     }
     \ENDIF
  }
    \ENDWHILE
 \STATE Answer ``Commutative''
\end{algorithmic}
\end{algorithm}

The idea is to form subsets of the set of matrices, matrix, and
row/column and query all the entries corresponding to them at a
setup stage. At each step of the walk, we swap a new set, a new
matrix, and a new row/column and update the entries accordingly.
The checking is done by computing the product of each pairs of
matrices without any further query. See Algorithm~\ref{alg:5} for
details. Then, the setup cost is $O(trsn)$, because we need to
query $s$ rows for each of $r$ matrices in each of $t$ sets. The
update cost is $O(rsn+tsn+rtn)$, $rsn$ for when swapping sets,
$tsn$ for when swapping matrices, and $O(rtn)$ for when swapping
rows/columns, \eg\ for a new set, we need to query the entries of
$r$ matrices, and for each matrix, we keep $s$ rows and columns.
The checking cost is $0$ because we have already queried the
entries of submatrices at the setup and the updating stages. The
eigenvalue gap, $\delta=\min\{1/t, 1/r, 1/s\}$, and $\epsilon
\approx \frac{r^2s^2t^2}{k^2n^2m^2}$ for $t\in o(m)$, $r \in
o(k)$, and $s \in o(n)$. Then our query complexity is
\[
trsn + \frac{1}{\sqrt{\delta \epsilon}}(rsn+tsn+rtn)
\]
for $\delta$, $\epsilon$ as stated above. By optimizing this, we
get a cost of $O(m^{6/7}k^{6/7}n^{13/7})$ with
$t=r=s=m^{2/7}k^{2/7}n^{2/7}$ for $k^{5/2}n^{5/2}\leq m$,
$m^{5/2}n^{5/2}\leq k$, and $m^{5/2}k^{5/2}\leq n$. $O(kmn^2)$
otherwise. On the other hand, if we perform a simple Grover's
search by searching on a pair of sets and within each pair of set,
a pair of noncommutative matrices, then it costs $O(mkn^{5/3})$.
Applying element distinctness over pairs of sets and within each
pair, applying Grover's search over $O(k^2)$ pairs of matrices,
and for each pair of matrices, applying a single pair
commutativity testing algorithm in Section~\ref{sec:MTSinglePair}
gives $O(m^{2/3}kn^{5/3})$ query complexity.

\subsection{Lower Bound}\label{commutativity-general-ex-lb}
$\Omega(m^{1/2}k^{1/2}n)$ lower bound is obtained by quantum
adversary argument.

Let $A$ be the set such that $m/2$ sets contain pseudo-identity
matrices, \ie\ for $1 \leq i \leq m/2$, $1 \leq j \leq k$, the
$j$-th matrix in $i$-th set consists of diagonal entries of all
$ij$. The other $m/2$ sets contain matrices with all the entries
being the same and non-zero. For $m/2 < i \leq m$, the $j$-th
matrix in $i$-th set contains all $ij$ entries. Then within each
of the $m$ sets, the matrices commute with each other. Also all of
$mk$ matrices commute with each other.

Let $B$ be the set such that one of $k$ matrices in one of $m/2$
sets that contain pseudo-identity matrices has one of off diagonal
entries being flipped from zero to the same entry as in diagonal.
Then within this set, the matrices still commute with each other
because the rest of the $k-1$ matrices are pseudo-identity. Within
each of the other sets, the matrices still commute, because they
are not affected. However, a set consists of the matrices from the
modified set and the matrices from one of $m/2$ sets that contain
all-same-entry matrices, gives non-commutative pairs. $m=m/2kn^2$,
$m'=1$ and $l=1$. So the lower bound is
$\sqrt{m/2kn^2}=\Omega(m^{1/2}k^{1/2}n)$.

\chapter{Summary and Future Work}\label{chap:conclusions}
We have seen two different kinds of quantum walk; Ambainis-Walk
and Szegedy-Walk, which are tools for providing upper bounds for
triangle finding problem and other matrix related problems. Both
of the walks give the same query upper bound for matrix product
verification. However, for triangle finding problem, Ambainis-Walk
gives a better query upper bound. In fact, we have shown that with
the same setup, update and checking cost for time or query
complexity, Ambainis-Walk gives a better bound. On the other hand,
Szegedy-Walk gives a better upper bound for time complexity in
matrix verification problem. Moreover, there is an algorithm for
testing commutativity of a general group~\cite{MN05}, where
analysis of Szegedy-walk is more powerful.

Both of these walks are \emph{discrete} in the sense that each
time step of the walk is discrete. There is another kind of walk
called \emph{continuous} walk, where the walk is performed with a
time step $\epsilon$ where $\epsilon\rightarrow 0$. There is an
application of continuous walk that gives an exponential
separation in quantum query complexity~\cite{CCD+03} from the
classical counterpart. There is no exponential separation shown
using discrete time walk so far, however. For some problem such as
a search on $N \times N$ grid, discrete walk performs
quadratically better than continuous walk without
ancilla~\cite{AKR05}. Whether discrete walk is more powerful than
continuous walk is an open question, although it is suspected that
these give essentially the same behaviour.

We have also seen Ambainis's quantum adversary theorem for proving
lower bounds. This technique is used to prove a lower bound of
$\Omega(\sqrt{n})$ for a search on unstructured database. From
this problem, we may derive lower bounds for many of the problems
studied in this essay.

For testing the commutativity of $k$ matrices of size $n \times
n$, we learned that there are three query complexities
$O(kn^{5/3})$, $O(k^{2/3}n^2)$ and $O(k^{4/5}n^{9/5})$ and
depending on the relationship between $k$ and $n$, one upper bound
is better than the others. The lower bound for this problem is
$\Omega(k^{1/2}n)$.

For future work, we would like to classify what kinds of problems
are better suited using Ambainis or Szegedy Walk. Also, we would
like to come up with an upper bound for the matrix commutativity
testing problem, that either supersedes or incorporates all the
three upper bounds. Since the gap between the current upper bound
and the lower bound is wide, we need to close the gap as well. We
are not sure if quantum adversary method can prove a tight lower
bound for this problem, and investigating other lower bound
methods is also of interest.


\bibliographystyle{alpha}
\bibliography{essay}
\end{document}